\documentclass[twocolumn,tighten]{aastex63}
\usepackage{color}
\usepackage{xspace}

\newcommand{\HI}{{H\,{\sc i}}\xspace}

\usepackage{multirow}

\shortauthors{Sano, Plucinsky, Bamba et al.}
\shorttitle{ALMA CO Observations of Gamma-Ray SNR N132D in the LMC}

\received{2020 July 14}
\revised{2020 August 30}
\accepted{2020 August 31}

\begin{document}
\title{ALMA CO Observations of Gamma-Ray Supernova Remnant N132D in the Large Magellanic Cloud:\\Possible Evidence for Shocked Molecular Clouds Illuminated by Cosmic-Ray Protons}

\author[0000-0003-2062-5692]{H. Sano}
\affiliation{National Astronomical Observatory of Japan, Mitaka, Tokyo 181-8588, Japan; hidetoshi.sano@nao.ac.jp}
\affiliation{Department of Physics, Nagoya University, Furo-cho, Chikusa-ku, Nagoya 464-8601, Japan}
\affiliation{Institute for Advanced Research, Nagoya University, Furo-cho, Chikusa-ku, Nagoya 464-8601, Japan}

\author[0000-0003-1415-5823]{P. P. Plucinsky}
\affiliation{Center for Astrophysics $|$ Harvard \& Smithsonian, 60 Garden St., Cambridge, MA 02138, USA}

\author[0000-0003-0890-4920]{A. Bamba}
\affiliation{Department of Physics, Graduate School of Science, The University of Tokyo, 7-3-1 Hongo, Bunkyo-ku, Tokyo 113-0033, Japan}
\affiliation{Research Center for the Early Universe, School of Science, The University of Tokyo, 7-3-1 Hongo, Bunkyo-ku, Tokyo 113-0033, Japan}

\author[0000-0003-3347-7094]{P. Sharda}
\affiliation{Research School of Astronomy and Astrophysics, Australian National University, Canberra, ACT 2611, Australia}
\affiliation{Australian Research Council Centre of Excellence for All Sky Astrophysics in 3 Dimensions (ASTRO 3D), Australia}

\author[0000-0002-4990-9288]{M. D. Filipovi{\'c}}
\affiliation{Western Sydney University, Locked Bag 1797, Penrith South DC, NSW 1797, Australia}

\author[0000-0003-1413-1776]{C. J. Law}
\affiliation{Center for Astrophysics $|$ Harvard \& Smithsonian, 60 Garden St., Cambridge, MA 02138, USA}

\author[0000-0001-5609-7372]{R. Z. E. Alsaberi}
\affiliation{Western Sydney University, Locked Bag 1797, Penrith South DC, NSW 1797, Australia}

\author[0000-0001-8296-7482]{Y. Yamane}
\affiliation{Department of Physics, Nagoya University, Furo-cho, Chikusa-ku, Nagoya 464-8601, Japan}

\author[0000-0002-2062-1600]{K. Tokuda}
\affiliation{National Astronomical Observatory of Japan, Mitaka, Tokyo 181-8588, Japan; hidetoshi.sano@nao.ac.jp}
\affiliation{Department of Physical Science, Graduate School of Science, Osaka Prefecture University, 1-1 Gakuen-cho, Naka-ku, Sakai 599-8531, Japan}

\author[0000-0002-6606-2816]{F. Acero}
\affiliation{Service d’Astrophysique, CEA Saclay, F-91191 Gif-sur-Yvette Cedex, France}

\author[0000-0001-5302-1866]{M. Sasaki}
\affiliation{Dr. Karl Remeis-Sternwarte, Erlangen Centre for Astroparticle Physics, Friedrich-Alexander-Universit$\ddot{a}$t Erlangen-N$\ddot{u}$rnberg, Sternwartstra$\beta$e 7, D-96049 Bamberg, Germany}

\author[0000-0002-4708-4219]{J. Vink}
\affiliation{Anton Pannekoek Institute/GRAPPA, University of Amsterdam, P.O. Box 94249, 1090 GE Amsterdam, The Netherlands}

\author{T. Inoue}
\affiliation{Department of Physics, Nagoya University, Furo-cho, Chikusa-ku, Nagoya 464-8601, Japan}

\author[0000-0003-4366-6518]{S. Inutsuka}
\affiliation{Department of Physics, Nagoya University, Furo-cho, Chikusa-ku, Nagoya 464-8601, Japan}

\author[0000-0003-3383-2279]{J. Shimoda}
\affiliation{Department of Physics, Nagoya University, Furo-cho, Chikusa-ku, Nagoya 464-8601, Japan}

\author[0000-0002-2794-4840]{K. Tsuge}
\affiliation{Department of Physics, Nagoya University, Furo-cho, Chikusa-ku, Nagoya 464-8601, Japan}

\author{K. Fujii}
\affiliation{Department of Physics, Graduate School of Science, The University of Tokyo, 7-3-1 Hongo, Bunkyo-ku, Tokyo 113-0033, Japan}

\author{F. Voisin}
\affiliation{School of Physical Sciences, The University of Adelaide, North Terrace, Adelaide, SA 5005, Australia}

\author[0000-0003-2762-8378]{N. Maxted}
\affiliation{School of Science, University of New South Wales, Australian Defence Force Academy, Canberra, ACT 2600, Australia}

\author[0000-0002-9516-1581]{G. Rowell}
\affiliation{School of Physical Sciences, The University of Adelaide, North Terrace, Adelaide, SA 5005, Australia}

\author[0000-0001-7826-3837]{T. Onishi}
\affiliation{Department of Physical Science, Graduate School of Science, Osaka Prefecture University, 1-1 Gakuen-cho, Naka-ku, Sakai 599-8531, Japan}

\author[0000-0001-7813-0380]{A. Kawamura}
\affiliation{National Astronomical Observatory of Japan, Mitaka, Tokyo 181-8588, Japan; hidetoshi.sano@nao.ac.jp}

\author{N. Mizuno}
\affiliation{National Astronomical Observatory of Japan, Mitaka, Tokyo 181-8588, Japan; hidetoshi.sano@nao.ac.jp}

\author{H. Yamamoto}
\affiliation{Department of Physics, Nagoya University, Furo-cho, Chikusa-ku, Nagoya 464-8601, Japan}

\author[0000-0002-1411-5410]{K. Tachihara}
\affiliation{Department of Physics, Nagoya University, Furo-cho, Chikusa-ku, Nagoya 464-8601, Japan}

\author{Y. Fukui}
\affiliation{Department of Physics, Nagoya University, Furo-cho, Chikusa-ku, Nagoya 464-8601, Japan}
\affiliation{Institute for Advanced Research, Nagoya University, Furo-cho, Chikusa-ku, Nagoya 464-8601, Japan}

\begin{abstract}
N132D is the brightest gamma-ray supernova remnant (SNR) in the Large Magellanic Cloud (LMC). We carried out $^{12}$CO($J$ = 1--0, 3--2) observations toward the SNR using the Atacama Large Millimeter/submillimeter Array (ALMA) and Atacama Submillimeter Telescope Experiment. We find diffuse CO emission not only at the southern edge of the SNR as previously known, but also inside the X-ray shell. We spatially resolved nine molecular clouds using ALMA with an angular resolution of $5''$, corresponding to a spatial resolution of $\sim$1 pc at the distance of the LMC. Typical cloud sizes and masses are $\sim$2.0 pc and $\sim$100 $M_\odot$, respectively. High-intensity ratios of CO $J$ = 3--2 / 1--0 $> 1.5$ are seen toward the molecular clouds, indicating that shock-heating has occurred. Spatially resolved X-ray spectroscopy reveals that thermal X-rays in the center of N132D are produced not only behind a molecular cloud, but also in front of it. Considering the absence of a thermal component associated with the forward shock towards one molecular cloud located along the line of sight to the center of the remnant, this suggests that this particular cloud is engulfed by shock waves and is positioned on the near side of remnant. If the hadronic process is the dominant contributor to the gamma-ray emission, the shock-engulfed clouds play a role as targets for cosmic-rays. We estimate the total energy of cosmic-ray protons accelerated in N132D to be $\sim$0.5--$3.8 \times 10^{49}$ erg as a conservative lower limit, which is similar to that observed in Galactic gamma-ray SNRs.

\end{abstract}

\section{Introduction}
It has been a long-standing question how cosmic rays, consisting of mainly relativistic protons, are accelerated in interstellar space. Supernova remnants (SNRs) are promising candidates for acceleration sites of Galactic cosmic rays below the knee energy ($\sim$3$ \times 10^{15}$~eV), through the mechanism of diffusive shock acceleration \citep[DSA, e.g.,][]{1978MNRAS.182..147B,1978ApJ...221L..29B} at their shocks. A conventional value of the total energy of Galactic cosmic-rays accelerated in a SNR is thought to be $\sim10^{49}$--$10^{50}$~erg, corresponding to $\sim$1--10\% of the typical kinematic energy released by a supernova explosion \citep[10$^{51}$~erg; e.g.,][]{2019AJ....158..149L}. One of the current challenges is to verify these predictions experimentally.

A young SNR (a few thousand years old) with bright TeV gamma-ray emission is a potential source for accelerating cosmic rays close to knee energy \citep[c.f.,][]{2012MNRAS.427...91O,2015ARNPS..65..245F,2018SSRv..214...41B}. TeV gamma-rays from young SNRs can be generally produced by two different mechanisms: hadronic and leptonic processes \citep[e.g.,][]{1994A&A...285..645A,1994A&A...287..959D,2010ApJ...708..965Z}. For the hadronic process, interaction between cosmic-ray proton and interstellar proton creates a neutral pion that decays into two gamma-ray photons (refer to as ``hadronic gamma-rays''). For the leptonic process, cosmic-ray electron energizes an interstellar photon to gamma-ray energy via inverse Compton scattering (refer to as ``leptonic gamma-rays''). To establish the SNR origin of cosmic-ray protons, an observational detection of hadronic gamma-ray is needed. However, it is difficult to distinguish the hadronic/leptonic processes from spectral modeling alone \citep[e.g.,][]{2012ApJ...744...71I,2018AA...612A...4H,2018AA...612A...6H,2018AA...612A...7H}.

Investigating the interstellar gas associated with gamma-ray SNRs holds a key to solving this problem. If the hadronic process dominates, the gamma-ray flux is proportional to the number density of the interstellar gas assuming azimuthally-isotropic distribution of cosmic rays. This implies that the presence of gamma rays should be spatially coincident with the interstellar gas. \cite{2012ApJ...746...82F} demonstrated the spatial correspondence using CO/\HI\ datasets as interstellar molecular and atomic gas tracers and TeV gamma-ray data toward the Galactic SNR RX~J1713.7$-$3946. The authors also derived the total energy of cosmic rays to be $\sim$10$^{48}$ erg, by adopting the number density of interstellar gas that interacts with the SNR. Subsequent studies toward the young TeV gamma-ray SNRs HESS~J1731$-$347, Vela Jr., and RCW~86 in the Milky Way show similar values of $\sim10^{48}$--$10^{49}$~erg \citep[e.g.,][]{2014ApJ...788...94F,2017ApJ...850...71F,2019ApJ...876...37S}. To better understand the origin of cosmic rays and their energy budget, we need to study not only Galactic SNRs, but also extragalactic sources such as gamma-ray bright SNRs in the nearby Large Magellanic Cloud (LMC) and the Small Magellanic Cloud (SMC).

\section{Overview of the Magellanic SNR N132D}
N132D (LHA\,120-N\,132D) is the brightest X-ray and TeV gamma-ray SNR in the LMC \citep[e.g.,][]{2016A&A...585A.162M,2015Sci...347..406H,2019scta.book..125M}. The shell-like morphology of this remnant is clearly resolved in radio, infrared, optical, and X-ray wavelengths \citep[e.g.,][]{1995AJ....109..200D,1996AJ....112..509M,2006ApJ...653..267T,2007ApJ...671L..45B}, with a size of $114'' \times 90''$\footnote{See also, catalog papers of the LMC SNRs by \cite{2010MNRAS.407.1301B} and \cite{2017ApJS..230....2B}.} or $\sim$25 pc at the distance of the LMC \citep[50 $\pm$ 1.3 kpc,][]{2013Natur.495...76P}. The age of the SNR is estimated to be $\sim$2500 yr \citep{1995AJ....109.2104M,1998ApJ...505..732H,2011Ap&SS.331..521V,2020ApJ...894...73L}. N132D is also categorized as an Oxygen-rich (or O-rich) SNR, which most likely originated from a core-collapse supernova explosion \citep[e.g.,][]{1976PASP...88...44D,1978ApJ...223..109L,1987ApJ...314..103H,2000ApJ...537..667B,Sharda2020}. Detailed optical studies revealed kinematic motions of O-rich ejecta using Doppler reconstructions with an average expansion velocity 1745~km~s$^{-1}$ \citep[][]{1980ApJ...237..765L,1995AJ....109.2104M,2009ApJ...707L..27F,2011Ap&SS.331..521V,2020ApJ...894...73L}.

Since the detection of TeV gamma ray emission associated with N132D, it has received much attention as a possible efficient accelerator of cosmic rays. The \cite{2015Sci...347..406H} first reported the significant detection of TeV gamma-rays toward three sources in the LMC, including the superbubble 30~Doradus~C, and two SNRs N157B and N132D. The authors derived the 1--10 TeV gamma-ray luminosity of $(0.9 \pm 0.2) \times 10^{35}$ erg s$^{-1}$ for N132D at the assumed distance of 50 kpc, which is an order of magnitude higher than that of the young ($\sim$1600 yr) TeV gamma-ray SNR RX~J1713.7$-$3946 in the Galactic plane. A subsequent GeV gamma-ray study using {\it{Fermi}}-LAT reported a 1--100 GeV gamma-ray luminosity of $\sim$10$^{36}$ erg s$^{-1}$, indicating that N132D is the brightest GeV gamma-ray SNR not only in the Magellanic Clouds, but also in the Local Group galaxies \citep{2016A&A...586A..71A,2016ApJS..224....8A}. \cite{2018ApJ...854...71B} {discovered hard X-ray emission ($E$: 10--15~keV)} using {\it{NuSTAR}}. {The authors} derived an upper limit on the synchrotron X-ray flux of $2.0 \times 10^{35}$~erg~s$^{-1}$ in the 2--10 keV band using {\it{Suzaku}} and {\it{NuSTAR}}{, and} argued that a high flux ratio of TeV gamma-ray and synchrotron X-rays is consistent with the hadronic origin of gamma-rays. However, to estimate the total energy of cosmic rays, the number density of interacting molecular and atomic clouds is needed.

N132D is also believed to be associated with a giant molecular cloud (GMC) that might be a possible target for cosmic-ray protons. \cite{1997ApJ...480..607B} discovered a GMC toward the south of N132D using $^{12}$CO($J$ = 2--1) line emission with the Swedish-ESO Submillimeter Telescope (SEST). The GMC has a size of $\sim$22~pc and a virial mass of at least $\sim2 \times 10^5$ $M_\odot$. The authors suggested that a part of the GMC is interacting with the southern edge of the SNR. This interpretation was further supported by the presence of shock-heated dust components in the southeastern shell of N132D \citep[e.g.,][]{2006ApJ...652L..33W,2006ApJ...653..267T,2012ApJ...754..132T,2013ApJ...779..134S,2018ApJS...237..10D,2019ApJ...882..135Z}. Subsequently, \cite{2010AJ....140..584D} and \cite{2015ASPC..499..257S} presented a CO map using archival $^{12}$CO($J$ = 1--0) line emission data which was taken by the Mopra 22-m radio telescope as part of the Magellanic Mopra Assessment project \citep[MAGMA,][]{2011ApJS..197...16W}. A diffuse part of the GMC is possibly aligned with the southern shell of the SNR, while no dense clouds are found inside the shell. Owing to the modest angular resolution of the CO data of $\sim$23$''$--$45''$ (or $\sim$6--11~pc at the LMC distance) and lack of higher excitation line data (e.g., $^{12}$CO $J$ = 3--2, 4--3), there is no conclusive evidence for shock-heated molecular clouds in the existing data of this remnant.

In this study, we report new millimeter/submillimeter observations using $^{12}$CO($J$ = 1--0, 3--2) line emission with the Atacama Submillimeter Telescope Experiment (ASTE) and the Atacama Large Millimeter/submillimeter Array (ALMA). The high angular resolution of 5$''$ in the {ALMA} CO data will allow us to resolve molecular clouds illuminated by shock waves and cosmic-ray protons in N132D. Section \ref{ss:obs} gives details about the observations, data reductions, and archival data. Sections \ref{ss:large} and \ref{ss:aste} show a large-scale view of the CO, \HI, X-ray, and TeV gamma-ray emission; Section~\ref{ss:alma} presents ALMA CO results and basic properties of the molecular clouds; Section \ref{ss:ratio} discusses the excitation condition of the molecular clouds; Section \ref{ss:orich} gives a detailed comparison with the O-rich ejecta; Sections \ref{ss:xspec} and \ref{ss:HardXray} present X-ray spectroscopy and a comparison with hard X-ray emission. Discussion and conclusions are given in Sections \ref{s:discussion} and \ref{s:conclusions}, respectively.

\section{Observations, Data Reductions, and Archival Data}\label{ss:obs}
\subsection{CO}
Observations of $^{12}$CO($J$ = 3--2) line emission at $\lambda$=0.87~mm wavelength were conducted in 2014 September 1--3 (PI: H. Sano, proposal\# AC141006) using the ASTE 10-m radio telescope \citep{2004SPIE.5489..763E}. The telescope is installed at an altitude of 5000~m in the Atacama Desert in Chile, operated by the Chile Observatory of the National Astronomical Observatory of Japan (NAOJ). We used the on-the-fly mapping mode with Nyquist sampling, and the effective observation area was $3.6' \times 3.6'$ centered at ($\alpha_\mathrm{J2000}$, $\delta_\mathrm{J2000}$) $\sim$ ($05^\mathrm{h}25^\mathrm{m}2$\farcs$8$, $-69\degr38\arcmin35\arcsec$). The front end was a sideband-separating Superconductor-Insulator-Superconductor (SIS) mixer receiver ``CATS 345'' \citep{2008stt..conf..281I}. We utilized an XF-type digital spectrometer ``MAC'' \citep{2000SPIE.4015...86S} as the backend. The bandwidth of MAC is 128~MHz with 1024 channels, corresponding to a spectral resolution of 0.125~MHz. The velocity coverage and resolution are thus $\sim$111~km~s$^{-1}$ and $\sim$0.11~km~s$^{-1}$, respectively. The typical system temperature was $\sim$300~K, including the atmosphere in the single-side band. To derive the main beam efficiency, we observed N\,159W \citep[$\alpha_\mathrm{J2000}, \delta_\mathrm{J2000}$=$05^\mathrm{h}40^\mathrm{m}3$\farcs$7$, $-68\degr47\arcmin00\arcsec$;][]{2011AJ....141...73M}, obtaining a main beam efficiency of 0.67$\pm$0.08. We also observed the M-type AGB star R~Dor every hour to satisfy pointing offsets accuracy within $2''$. After convolution with a two-dimensional Gaussian kernel, we obtained the cube data with the beam size of $\sim$$23''$ ($\sim$5.6~pc at the LMC distance). The typical noise fluctuations are $\sim$0.046~K at the velocity resolution of 0.4~km~s$^{-1}$.

Observations of $^{12}$CO($J$ = 1--0) line emission at $\lambda$=2.6~mm wavelength were carried out using the ALMA Band~3 (86--116~GHz) in Cycle~2 as an early science project (PI: H. Sano, proposal\# 2013.1.01042.S). We utilized 40 antennas of 12-m array, 9 antennas of 7-m array, and 3 antennas of total power (TP) array. The effective observation area was a $150'' \times 150''$ rectangular region centered at ($\alpha_\mathrm{J2000}$, $\delta_\mathrm{J2000}$) $\sim$ ($05^\mathrm{h}25^\mathrm{m}2$\farcs$79$, $-69\degr38\arcmin34\farcs2$). The combined baseline length of 12-m and 7-m arrays ranges from 7.2 to 215.7~m, corresponding to {\it{u-v}} distances from 2.8 to 82.9~$k\lambda$. The two quasars J0334$-$4008 and J0635$-$7516 were used for complex gain calibrators. Another two quasars J0601$-$7036 and J0526$-$6749 were observed as phase calibrators. We also observed Callisto, Uranus, and a quasar J0519-454 as flux calibrators. The data reduction was performed using the Common Astronomy Software Application \citep[CASA;][]{2007ASPC..376..127M} package version 5.4.0. We used the {\it{multiscale CLEAN}} task implemented in the CASA package \citep{2008ISTSP...2..793C}. The scale parameters are $0''$, 3\farcs18, and 9\farcs54 for the 12-m array and $0''$, 14\farcs64, and 43\farcs92 for the 7-m array. To improve the imaging quality, we also applied a {\it{uvtaper}} during the clean procedure for the 12-m array data. The {\it{u-v}} tapering applies a multiplicative Gaussian taper to the spatial frequency space, to downweight high spatial frequencies. This can suppress artifacts --- e.g., strong side-lobes --- arising from poorly sampled areas near and beyond the maximum spatial frequency. We finally combined the cleaned data of 12- and 7-m array datasets and calibrated the TP array data by using the {\it{feather}} task. The final beam size of the feathered data is $5\farcs26 \times 4\farcs99$, with a position angle of $59\fdg60$, corresponding to a spatial resolution of $\sim$1.2~pc at the LMC distance. The typical noise fluctuations of the feathered data are $\sim$0.22~K at a velocity resolution of 0.4~km~s$^{-1}$.

To investigate the CO gas distribution at larger spatial scales, we used the Magellanic Mopra Assessment Data Release 1 \citep[MAGMA DR1,][]{2011ApJS..197...16W}. MAGMA is a $^{12}$CO($J$ = 1--0) mapping survey of the LMC using the Mopra 22-m radio telescope of the Australia Telescope National Facility (ATNF). The angular resolution is $45''$, corresponding to the spatial resolution of $\sim$11~pc at the LMC distance. The typical noise fluctuations of a region surrounding N132D are $\sim$0.26~K at the velocity resolution of 0.53~km~s$^{-1}$. We applied additional spatial smoothing with a two-dimensional Gaussian kernel. The angular resolution of smoothed data is $\sim$60$''$ ($\sim$15 pc at the LMC distance), which is the same resolution with the \HI\ survey data of the LMC (see Section \ref{ss:hi}).

\subsection{\rm \HI}\label{ss:hi}
To better understand the distribution of neutral atomic hydrogen toward N132D, we used an archival survey data of the \HI\ line at $\lambda$ = 21~cm wavelength published by \cite{2003ApJS..148..473K}. The survey data were obtained using the Australia Telescope Compact Array (ATCA) and Parkes 64-m telescopes operated by Australia Telescope National Facility (ATNF). The angular resolution of the survey data is $60''$, corresponding to the spatial resolution of $\sim$15~pc at the LMC distance. The typical noise fluctuations of brightness temperature are $\sim$2.4~K at the velocity resolution of 1.689~km~s$^{-1}$.

\subsection{X-rays}\label{ss:xray}

We used archival X-ray data obtained by {\it{Chandra}}, for which the observation IDs are 5532, 7259, and 7266 (PI: K. J. Borkowski, proposal\# 06500305), which have been published in previous papers \citep[e.g.,][]{2007ApJ...671L..45B,2008AdSpR..41..416X,2016AJ....151..161S,Sharda2020}. The datasets were taken with the Advanced CCD Imaging Spectrometer S-array (ACIS-S3). Table~\ref{tab:n132dlog} lists the details of the observations. We utilized Chandra Interactive Analysis of Observations \citep[CIAO,][]{2006SPIE.6270E..1VF} software version 4.12 with CALDB 4.9.1 \citep{2007ChNew..14...33G} for data reduction. All the datasets were reprocessed using the {\it{chandra\_repro}} task. We then created exposure-corrected, energy-filtered images using the {\it{fluximage}} task in the energy bands of 0.35--0.85~keV, 0.5--1.2~keV (soft-band), 0.85--1.6~keV, 1.2--2.0~keV (medium-band), 1.6--6.0~keV, 2.0--7.0~keV (hard-band), and 0.5--7.0~keV (broad-band). The total effective exposure is $\sim$89.3~ks. For the spectral analysis, we used HEASOFT (version 6.25), including spectral fitting with XSPEC (version 12.10.1f, \citealt{1996ASPC..101...17A}). We fit the spectra in the energy band 0.3--10.0~keV and the errors of the fitted parameters are quoted at the 1~$\sigma$ confidence level unless specified otherwise. We fit the unbinned spectra to preserve the maximum spectral information. Following \cite{Sharda2020}, we explicitly model the background as opposed to subtracting it. We use the C statistic \citep{Cash1979} as the minimization statistic to avoid the well-known bias introduced by the $\chi^2$ statistic in the case of a low number of counts per spectral bin \citep{Kaastra2017} and we report the Pearson $\chi^2$ (weighting by the model) to evaluate goodness of fit. We used ATOMDB version 3.0.9 \citep{2013Foster} and non-equilibrium ionization (NEI) version 3.0.4 for the NEI models \citep{2001ApJ...548..820B}. We used the cosmic abundance given by \cite{Wilms2000} as the baseline abundance for all our analysis and the cross sections given by \cite{Verner1996}. 

\begin{deluxetable}{l c c c c }
\tabletypesize{\scriptsize}
\label{tab:n132dlog}
\tablecolumns{6}
\tablecaption{{\it{Chandra}} ACIS-S observation log of SNR N132D.}
\tablehead{\colhead{ObsID} & \colhead{Observation Date} & \colhead{Exposure} & \colhead{$\alpha_{\mathrm{J2000}}$} & \colhead{$\delta_{\mathrm{J2000}}$}\\
&& (ks) & ($^{\mathrm{h}}$ $^{\mathrm{m}}$ $^{\mathrm{s}}$) & ($^{\circ}$ $\arcmin$ $\arcsec$)}
\startdata
05532&Jan 09, 2006&44.59 & 05 25 02.28 & $-$69 38 37.32\\
07259&Jan 10, 2006&24.85 & 05 25 02.28 & $-$69 38 37.32\\
07266&Jan 15, 2006&19.90 & 05 25 02.28 & $-$69 38 37.32\\
\enddata
\end{deluxetable}

To investigate the origin of the hard X-ray emission in N132D, we also used a map of the hard-band X-ray ($E$: 10--15 keV) obtained with {\it{NuSTAR}} \citep{2018ApJ...854...71B}. The {\it{NuSTAR}} observations were executed on 2015 December 10--11. {The total effective exposure is 62.3~ks. The angular resolution is $18''$ (half-power beam-width; HPBW) or $58''$ (half-power diameter; HPD).} To improve signal to noise ratios of the map, we smoothed the data with a two-dimensional Gaussian kernel of $20''$.

\subsection{TeV Gamma-Rays}
{To compare the spatial distribution with the interstellar medium (ISM) environment of N132D, we also used an excess count map of TeV gamma-rays obtained by the High Energy Stereoscopic System \citep{2015Sci...347..406H}}. The angular resolution is $\sim$$3'$ for the point spread function (PSF, 68\% containing radius) or $\sim$$7'$ for the full-width half-maximum (FWHM), corresponding to the spatial resolution of $\sim$44 pc for the PSF or $\sim$100 pc for FWHM.

\begin{figure*}[]
\begin{center}
\includegraphics[width=140mm,clip]{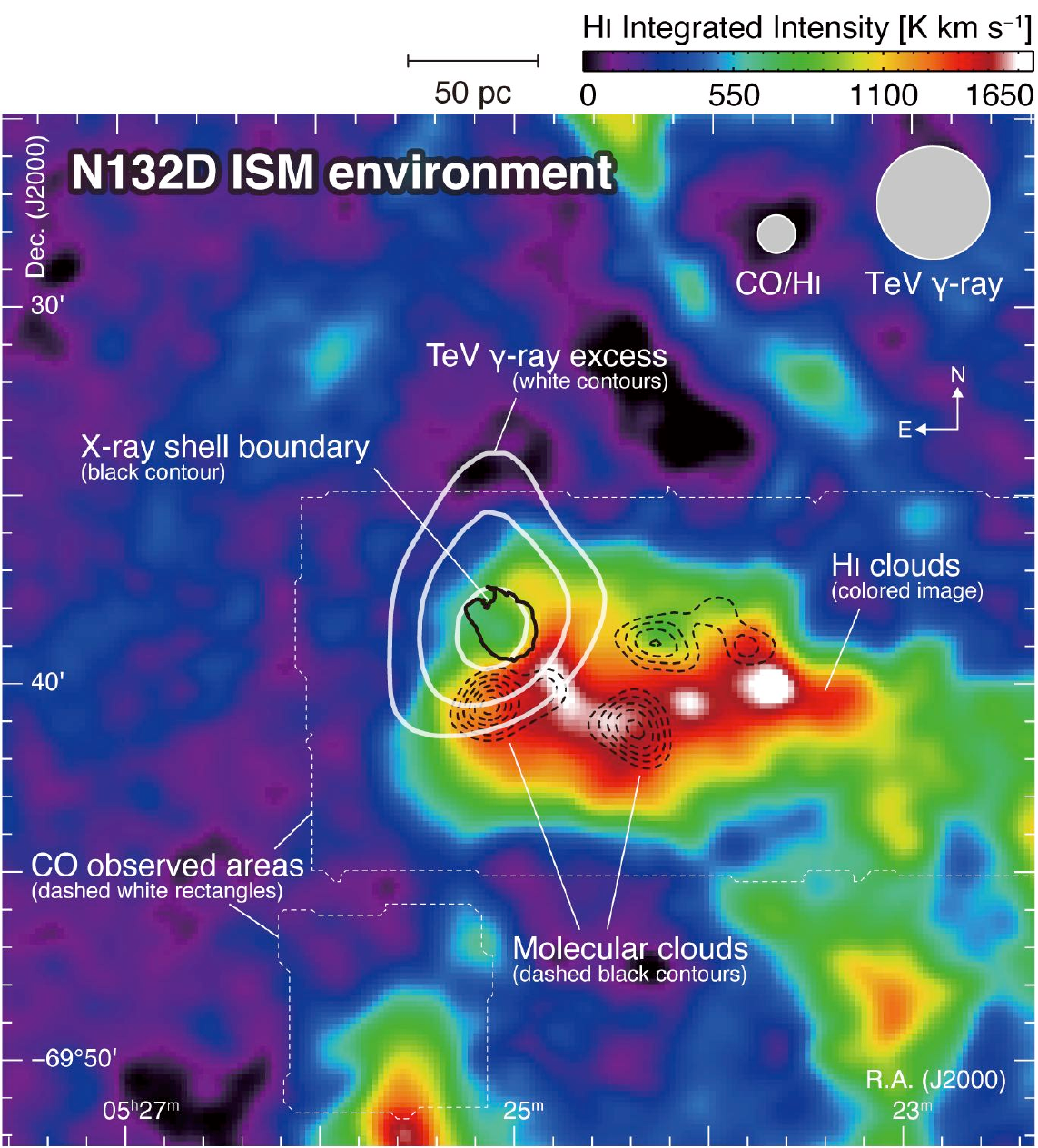}
\caption{Integrated intensity map of the ATCA \& Parkes \HI\ \citep[][$colored$ $image$]{2003ApJS..148..473K} overlaid with the Mopra $^{12}$CO($J$ = 1--0) integrated intensity ({\it black dashed contours}), {\it Chandra} X-ray boundary of N132D \citep[][$black$ $solid$ $contour$]{2007ApJ...671L..45B}, and the H.E.S.S. TeV gamma-ray excess counts \citep[][$white$ $solid$ $contours$]{2015Sci...347..406H}. The CO data have been spatially smoothed to match the FWHM of \HI\ ($\sim$$60''$). The integration velocity ranges of CO and \HI\ are from 240 to 290 km s$^{-1}$, which cover 74 \% of total integrated intensity of \HI. The lowest contour level and intervals of CO are {6.3} K km s$^{-1}$ ($\sim${9} $\sigma$) and 2.1 K km s$^{-1}$ ($\sim$3 $\sigma$), respectively. The contour level of X-ray boundary is $0.3 \times 10^{-6}$ counts pixel$^{-1}$ s$^{-1}$. The contour levels of TeV gamma-rays are 10, 14, and 18 excess counts. The dashed white rectangles indicate the observed areas of CO. The FWHM beam size of CO/\HI\ and the PSF of TeV gamma-rays are also shown in top right corner.} 
\label{fig1}
\end{center}
\end{figure*}%

\subsection{\rm H$\alpha$ and [O{\sc iii}]}
The optical data of H$\alpha$ and [O{\sc iii}] emission lines are used to derive the spatial distributions of the ionized gas and shocked ejecta. We utilized the {\it{Hubble Space Telescope (HST)}} Wide Field Planetary Camera 2 (WFPC2) and the Advanced Camera for Survey (ACS) images from the Hubble Legacy Archive\footnote{hla.stsci.edu}. The observations were carried out using the ACS Wide Field Channel F658N for H$\alpha$ and the WFPC2 F502N for [O{\sc iii}], which have been published by \cite{1996AJ....112..509M} and \cite{2007ApJ...671L..45B}. For further details about the data reductions and pipeline processes, we refer the reader to the HST Data Handbook\footnote{www.stsci.edu/hst/HST\_overview/documents/datahandbook}.


\section{Results}

\begin{figure*}[]
\begin{center}
\includegraphics[width=\linewidth,clip]{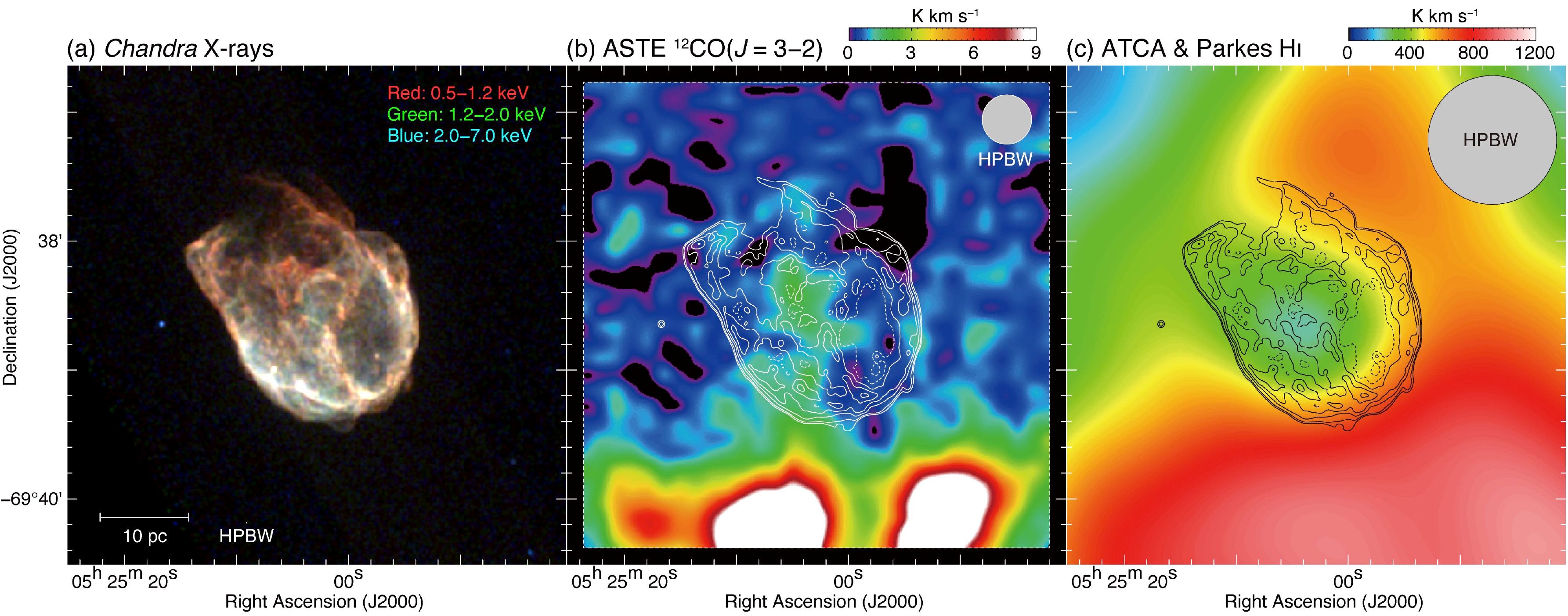}
 \caption{(a) RGB X-ray image of N132D obtained with {\it Chandra} \citep[e.g.,][]{2007ApJ...671L..45B,2008AdSpR..41..416X,Sharda2020}. The red, green, and blue colors represent the energy bands of 0.5--1.2~keV, 1.2--2.0~keV, and 2.0--7.0~keV, respectively. (b) Intensity map of $^{12}$CO($J$ = 3--2) obtained with ASTE overlaid with the {\it Chandra} X-ray contours. The integration velocity range is from 256.8~km~s$^{-1}$ to 271.2~km~s$^{-1}$. The black contours represent the X-ray {intensity}. The contour levels of X-rays are 0.3, 0.6, 1.8, 3.6, 7.2, and $14.4 \times 10^{-6}$~counts~pixel$^{-1}$~s$^{-1}$. (c) Intensity map of \HI\ obtained with ATCA and Parkes \citep{2003ApJS..148..473K} overlaid with the {\it Chandra} X-ray contours. The integration velocity range and contour levels are the same as in Figure~\ref{fig2}(b). The beam size and scale bar are also shown.}
\label{fig2}
\end{center}
\end{figure*}%

\subsection{Large-Scale Distribution of CO, {\rm \HI}, X-ray, and TeV Gamma-Rays}\label{ss:large}
Figure~\ref{fig1} shows a map of ATCA \& Parkes \HI\ intensity overlaid with the Mopra $^{12}$CO($J$ = 1--0) intensity (black dashed contours), {\it{Chandra}} X-ray boundary of N132D (black solid contours), and the H.E.S.S. TeV gamma-rays (white solid contours). An \HI\ cloud appears projected onto the SNR, which is elongated to the southwest direction with a hollow structure along the X-ray shell boundary. Three to four GMCs are located near the local intensity peaks of the \HI\ cloud. One of them is possibly associated with the southern shell boundary of the SNR, which is consistent with previous CO studies \citep[][]{1997ApJ...480..607B,2010AJ....140..584D,2015ASPC..499..257S}. Note that there are no dense molecular and atomic clouds toward Northeast outside the SNR. Further, note that TeV gamma-rays are emitted from the SNR itself rather than from the surrounding GMCs and \HI\ cloud, even after taking into consideration the large PSF of gamma-ray data.

\subsection{CO and {\rm \HI} Clouds toward the SNR}\label{ss:aste}
Figure~\ref{fig2}(a) shows an RGB image of N132D obtained with {\it{Chandra}}. The X-ray shell shows an incomplete elliptical morphology, slightly elongated in the southwestern direction, with a breakout structure in the northeast. Many filamentary structures of X-rays appear not only in the shell boundary, but also inside the SNR. The hard-band X-rays ($E$: 2.0--7.0~keV) are brighter in the southeastern shell.

Figures \ref{fig2}(b) and \ref{fig2}(c) show the integrated intensity maps of ASTE $^{12}$CO($J$ = 3--2) and ATCA \& Parkes \HI. Because of high sensitivity and full-spatial sampling observations of CO line emission, we found a molecular cloud toward the center of the SNR (hereafter ``N132D MC-center''). {Note that N132D MC-center is significantly detected because the CO integrated intensity of 1.12 K km s$^{-1}$ represents $\sim$10$\sigma$ level.} The spatially-resolved MC-center cloud is more extended than the beam size, with 10$\sigma$ or higher significance in integrated intensity. We also confirm the presence of the previously identified GMC (hereafter ``N132D GMC-south'') in contact with the southeastern edge of the SNR. The peak velocities of the clouds are $V_\mathrm{LSR} \sim$264~km~s$^{-1}$ for N132D MC-center and $V_\mathrm{LSR} \sim$266~km~s$^{-1}$ for N132D GMC-south, and the latter is roughly consistent with the previous CO observations using SEST \citep{1997ApJ...480..607B}. On the other hand, the overall distribution of \HI\ tends to encircle the X-ray shell except for northeast at the same velocity range of CO ($V_\mathrm{LSR} = 256.8$--271.2~km~s$^{-1}$). We also find that diffuse \HI\ gas with an intensity of $\sim$300~K~km~s$^{-1}$ fills the interior of the X-ray shell.

\begin{figure}[]
\begin{center}
\includegraphics[width=\linewidth,clip]{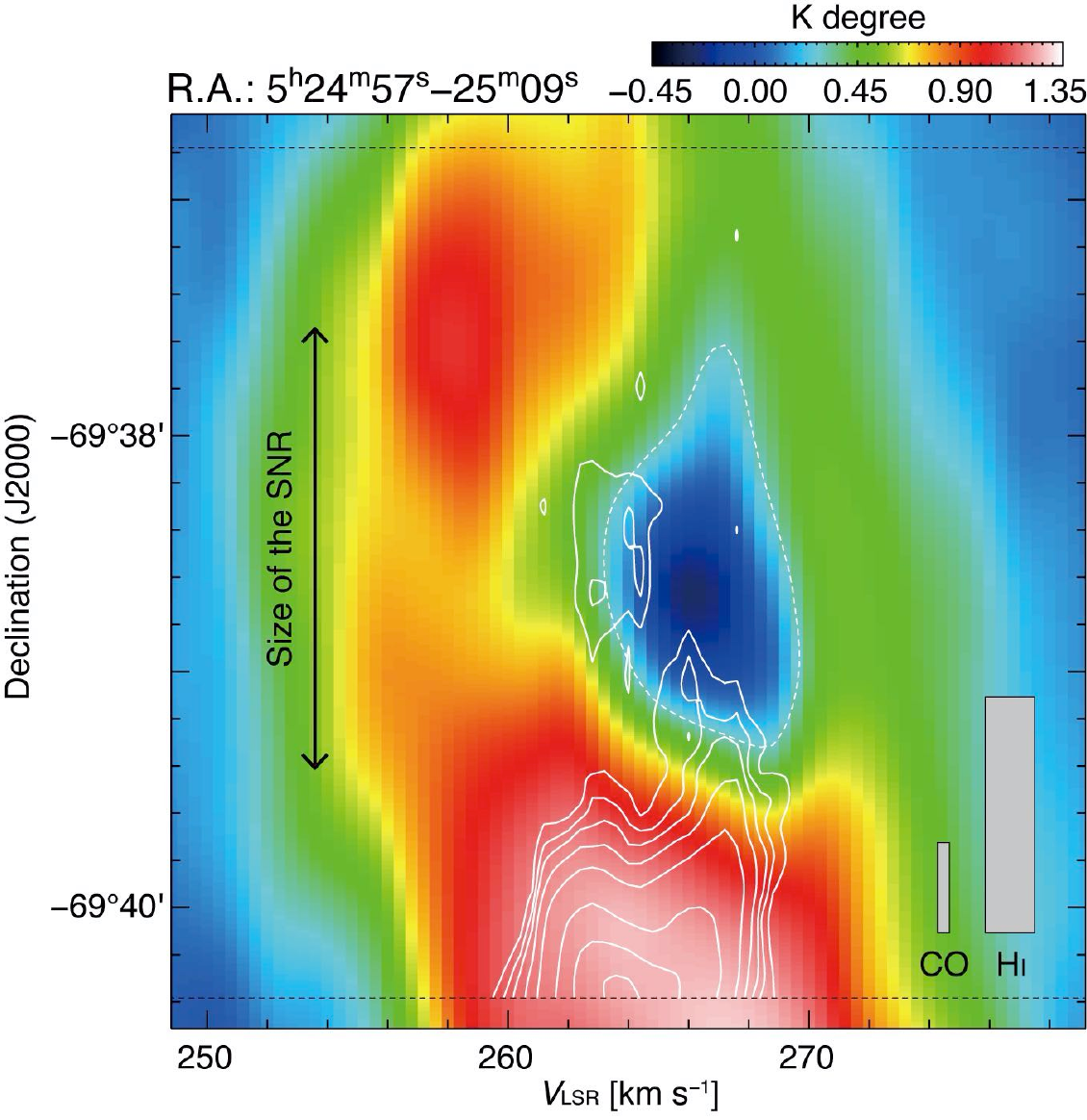}
\caption{Position-velocity diagram of {{H\,{\sc i}}\xspace}. Superposed white solid contours indicate $^{12}$CO($J$ = 3--2) intensity. The integration range in Right Ascension is from $05^\mathrm{h}24^\mathrm{m}57^\mathrm{s}$ to $05^\mathrm{h}25^\mathrm{m}09^\mathrm{s}$. The lowest contour and contour intervals of CO are 0.003 K degree and 0.005 K degree, respectively. The dashed white line delineates an \HI\ cavity with the \HI\ intensity of 0.3 K degree. The beam size and velocity resolution are also shown in bottom right corner. Horizontal dashed lines represent observed boundaries of CO.}
\label{fig3}
\end{center}
\end{figure}%

\begin{figure}[]
\begin{center}
\includegraphics[width=\linewidth,clip]{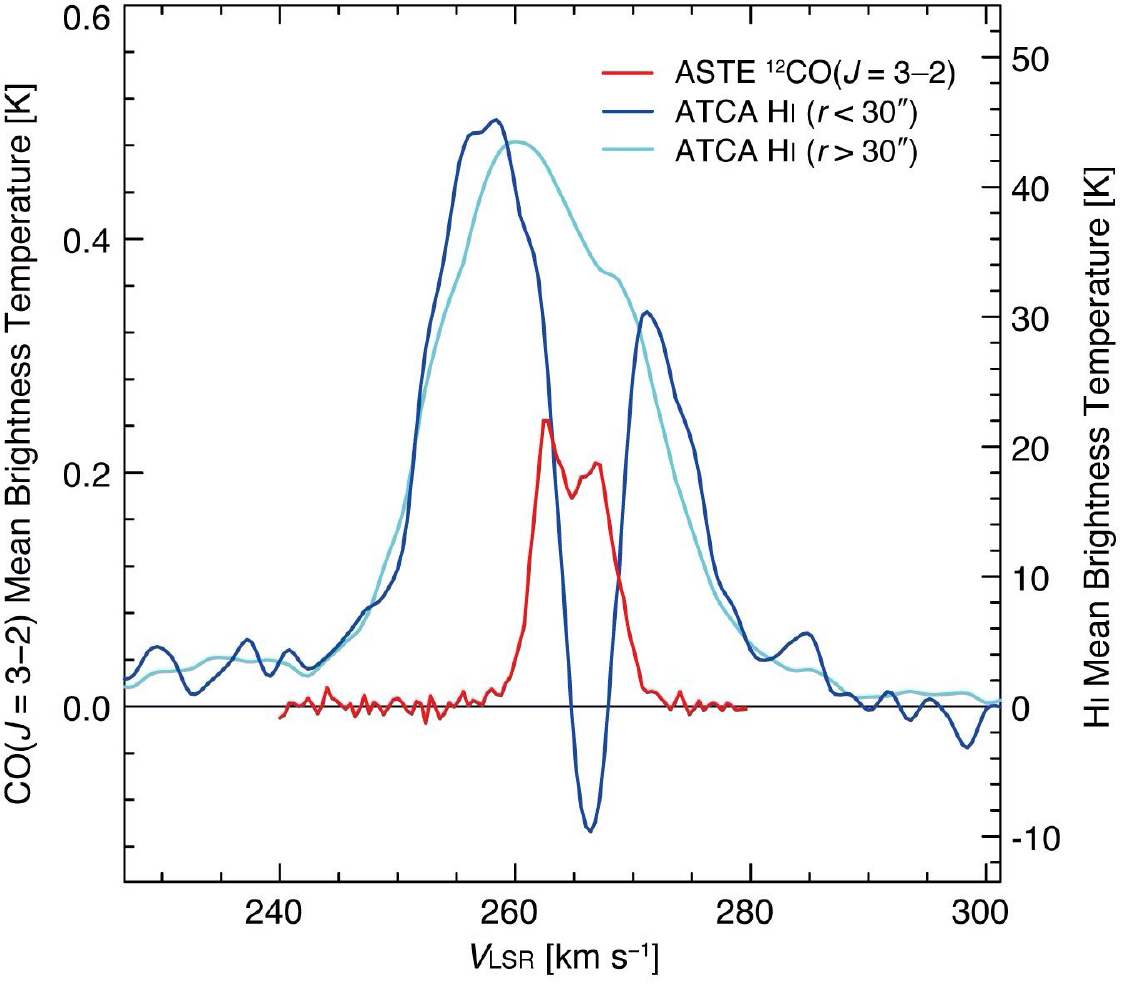}
\caption{Averaged line profiles of $^{12}$CO($J$ = 3--2) (red) and \HI\ (blue and cyan) toward the SNR N132D. The blue and cyan spectra were extracted from regions of the inside and outside of the circle with a radius of $30''$ at the center position of ($\alpha_\mathrm{J2000}$, $\delta_\mathrm{J2000}$) $\sim$ ($05^\mathrm{h}25^\mathrm{m}02\fs88$, $-69\degr38\arcmin34\farcs8$). The red spectrum was produced by averaging the whole spectra shown in Figure~\ref{fig2}b.}
\label{fig4}
\end{center}
\end{figure}%

Figure~\ref{fig3} shows a position-velocity diagram of CO and \HI. We find an intensity dip at the velocity of $\sim$266~km~s$^{-1}$, which is roughly centered at the position of the SNR in declination. On the other hand, the CO clouds appear projected onto the edge of the \HI\ dip at the intensity level of 0.3~K degree (dashed contour centered at $\sim$266~km~s$^{-1}$). Figure \ref{fig4} shows averaged line profiles of CO and \HI. The velocity range of \HI\ cloud at $V_\mathrm{LSR} \sim$250--280~km~s$^{-1}$ contains that of CO clouds at $V_\mathrm{LSR} \sim$260--270~km~s$^{-1}$. A strong absorption line of \HI\ is detected at the velocity of $V_\mathrm{LSR} \sim$266~km~s$^{-1}$ toward only the SNR direction (blue, inside the SNR), suggesting that the absorption line was caused by vicinity of strong radio continuum radiation from the SNR \citep[][]{2018ApJ...863...55Y,2018ApJ...867....7S,2019ApJ...873...40S}. We therefore focus on both the CO and \HI\ clouds around $V_\mathrm{LSR} \sim$266~km~s$^{-1}$ that are likely related with the SNR.

\begin{figure*}[]
\begin{center}
\includegraphics[width=145mm,clip]{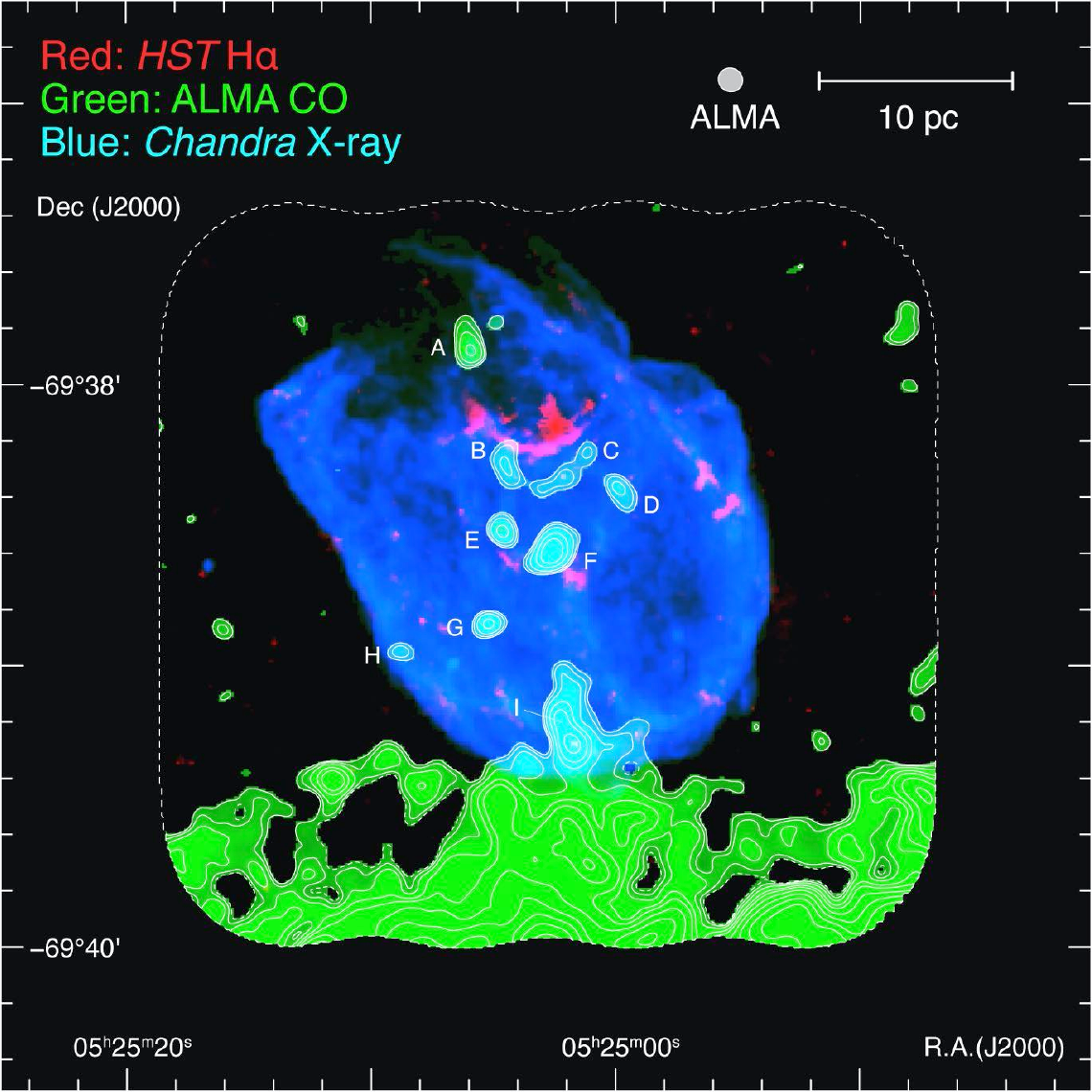}
\caption{RGB image of N132D obtained with the {\it{HST}} H$\alpha$ (red), ALMA $^{12}$CO($J$ = 1--0) (green), and {\it Chandra} X-rays in the energy band of 0.5--7.0~keV (blue). The integration velocity range is from 262.0~km~s$^{-1}$ to 268.4~km~s$^{-1}$. The contours represent the integrated intensity of CO, whose levels are 1.2 ($\sim3\sigma$), 1.5, 2.0, 3.0, 4.0, 6.0, 9.0, 13.0, 18.0, and 24.0~K~km~s$^{-1}$. The region enclosed by the dashed line indicates the observed area with ALMA. The CO clouds, named A--I, discussed in Section~\ref{ss:alma} are indicated.}
\label{fig5}
\end{center}
\end{figure*}%

\subsection{Detailed CO Distribution with ALMA}\label{ss:alma}

Figure~\ref{fig5} shows an RGB image of N132D composed of a combination of {\it{HST}} H$\alpha$ (red), ALMA $^{12}$CO($J$ = 1--0) integrated intensity (green), and the {\it{Chandra}} broad-band X-rays (blue). We spatially resolved nine molecular clouds, named A--I, within the X-ray shell of N132D. Cloud A is located in the breakout region with very faint X-rays. Clouds H and I lie on the edge of the southwestern shell. The other clouds, B--G, corresponding to N132D MC-center, are concentrated in the center of the SNR. In other words, N132D MC-center is split into clouds B--G owing to high-resolution observations using ALMA. Note that clouds B, C, E, and F are located in the vicinity of H$\alpha$ blobs or filaments as shown in red.

To derive the masses of these molecular clouds, we utilize the following equations:
\begin{eqnarray}
M = m_{\mathrm{p}} \mu \Omega D^2 \sum_{i} N_i(\mathrm{H}_2),\\
N(\mathrm{H}_2) = A \cdot W[\mathrm{^{12}CO}(J = 1\text{--}0)],
\label{eq1}
\end{eqnarray}
where $m_\mathrm{p}$ is the mass of atomic hydrogen, $\mu = 2.72$ is the mean molecular weight, $\Omega$ is the solid angle of each data pixel, $D$ is the distance to the LMC (= 50 kpc), $N_i(\mathrm{H}_2)$ is the column density of molecular hydrogen for each data pixel $i$, $A$ is the CO-to-H$_2$ conversion factor, and $W[\mathrm{^{12}CO}(J = 1\text{--}0)]$ is the integrated intensity of $^{12}$CO($J$ = 1--0) line emission. Here, we use the CO-to-H$_2$ conversion factor $A = 7.0 \times 10^{20}$~cm$^{-2}$ (K~km~s$^{-1})^{-1}$ \citep{2008ApJS..178...56F}. The size of each molecular cloud is defined as an effective diameter, determined by the contour of the half level of maximum integrated intensity. The detailed definitions and physical properties of molecular clouds are summarized in Table \ref{tab:mc}. The typical cloud masses and sizes are $\sim$50--100 $M_\sun $ and $\sim$1.5--2.0~pc, respectively. Note that the mass, $n$(H$_2$), and size of each cloud have $\sim$30 \% relative errors due to uncertainties in the CO-to-H$_2$ conversion factor and distance to the LMC. There are no broadline features with a velocity width more than 10~km~s$^{-1}$, whereas the linewidths of cloud A ($\Delta V$ = 5.7~km~s$^{-1}$) and cloud B ($\Delta V$ = 4.4~km~s$^{-1}$) are significantly larger than that of the other clouds ($\Delta V$ $\sim$1--2~km~s$^{-1}$).

\begin{deluxetable*}{lcccccccc}[]
\tablecaption{Physical properties of molecular clouds associated with N132D}
\tablehead{\\
\multicolumn{1}{c}{Cloud name} & $\alpha_{\mathrm{J2000}}$ & $\delta_{\mathrm{J2000}}$ & $T_{\mathrm{mb}} $ & $V_{\mathrm{LSR}}$ & $\Delta V$ & Size &  Mass & $n(\mathrm{H_2})$\\
& ($^{\mathrm{h}}$ $^{\mathrm{m}}$ $^{\mathrm{s}}$) & ($^{\circ}$ $\arcmin$ $\arcsec$) & (K) & \scalebox{0.9}[1]{(km $\mathrm{s^{-1}}$)} & \scalebox{0.9}[1]{(km $\mathrm{s^{-1}}$)} & (pc) &  ($M_\sun $) & (cm$^{-3}$)\\
\multicolumn{1}{c}{(1)} & (2) & (3) & (4) & (5) & (6) & (7) & (8) & (9)}
\startdata
A ................. & 05 25 05.94 & $-$69 37 52.9 & $0.79 \pm 0.08$ & $263.5 \pm 0.3$ & $5.7 \pm 0.7$ & 1.6 & \phantom{0}90 & 910\\
B ................. & 05 25 04.42 & $-$69 38 17.8 & $0.52 \pm 0.09$ & $263.5 \pm 0.4$ & $4.4 \pm 0.9$ & 1.8 & \phantom{0}70 & 450\\
C ................. & 05 25 01.17 & $-$69 38 14.1 & $0.74 \pm 0.18$ & $263.6 \pm 0.1$ & $1.1 \pm 0.3$ & 2.2 & \phantom{0}40 & 130\\
D ................. & 05 24 59.95 & $-$69 38 22.1 & $1.34 \pm 0.15$ & $262.6 \pm 0.1$ & $1.5 \pm 0.2$ & 1.7 & \phantom{0}50 &410\\
E ................. & 05 25 04.62 & $-$69 38 31.6 & $1.86 \pm 0.12$ & $264.1 \pm 0.1$ & $1.7 \pm 0.1$ & 1.4 & \phantom{0}60 & 790\\
F ................. & 05 25 02.69 & $-$69 38 35.8 & $2.79 \pm 0.16$ & $263.4 \pm 0.1$ & $1.7 \pm 0.1$ & 1.8 & 130 & 930\\
G ................. & 05 25 05.23 & $-$69 38 51.2 & $1.39 \pm 0.12$ & $264.7 \pm 0.1$ & $2.1 \pm 0.2$ & 1.2 & \phantom{0}40 & 820\\
H ................. & 05 25 08.58 & $-$69 38 57.0 & $1.66 \pm 0.17$ & $265.8 \pm 0.1$ & $0.9 \pm 0.1$ & 1.4 & \phantom{0}30 & 370\\
I .................. & 05 25 01.88 & $-$69 39 16.1 & $2.95 \pm 0.12$ & $266.7 \pm 0.1$ & $2.0 \pm 0.1$ & 2.1 & 240 & 990\\
\enddata
\tablecomments{Col. (1): Cloud name. Cols. (2--9): Observed physical properties of the clouds obtained by single or double Gaussian fitting with $^{12}$CO($J$ = 1--0) emission line. Cols. (2)--(3): Position of the clouds. Col. (4): Maximum radiation temperature. Col. (5): Central velocity of CO spectra. Col. (6): FWHM linewidth of CO spectra $\Delta V$. Col. (7): Diameter of clouds defined as $(S / \pi)^{0.5} \times 2$, where $S$ is the surface area of clouds surrounded by contours of the half level of maximum integrated intensity. Col. (8): Mass of clouds derived by an equation of $N(\mathrm{H_2}) / W(\mathrm{CO}) = 7.0 \times 10^{20}$ (K km s$^{-1}$)$^{-1}$ cm$^{-2}$, where $N(\mathrm{H_2})$ is the column density of molecular hydrogen and $W$(CO) is the integrated intensity of $^{12}$CO($J$ = 1--0) \citep{2008ApJS..178...56F}. Col. (9): Number density of molecular hydrogen $n(\mathrm{H_2})$}
\label{tab:mc}
\end{deluxetable*}


\subsection{CO 3--2 / 1--0 Ratio}\label{ss:ratio}

\begin{figure}[]
\begin{center}
\includegraphics[width=\linewidth,clip]{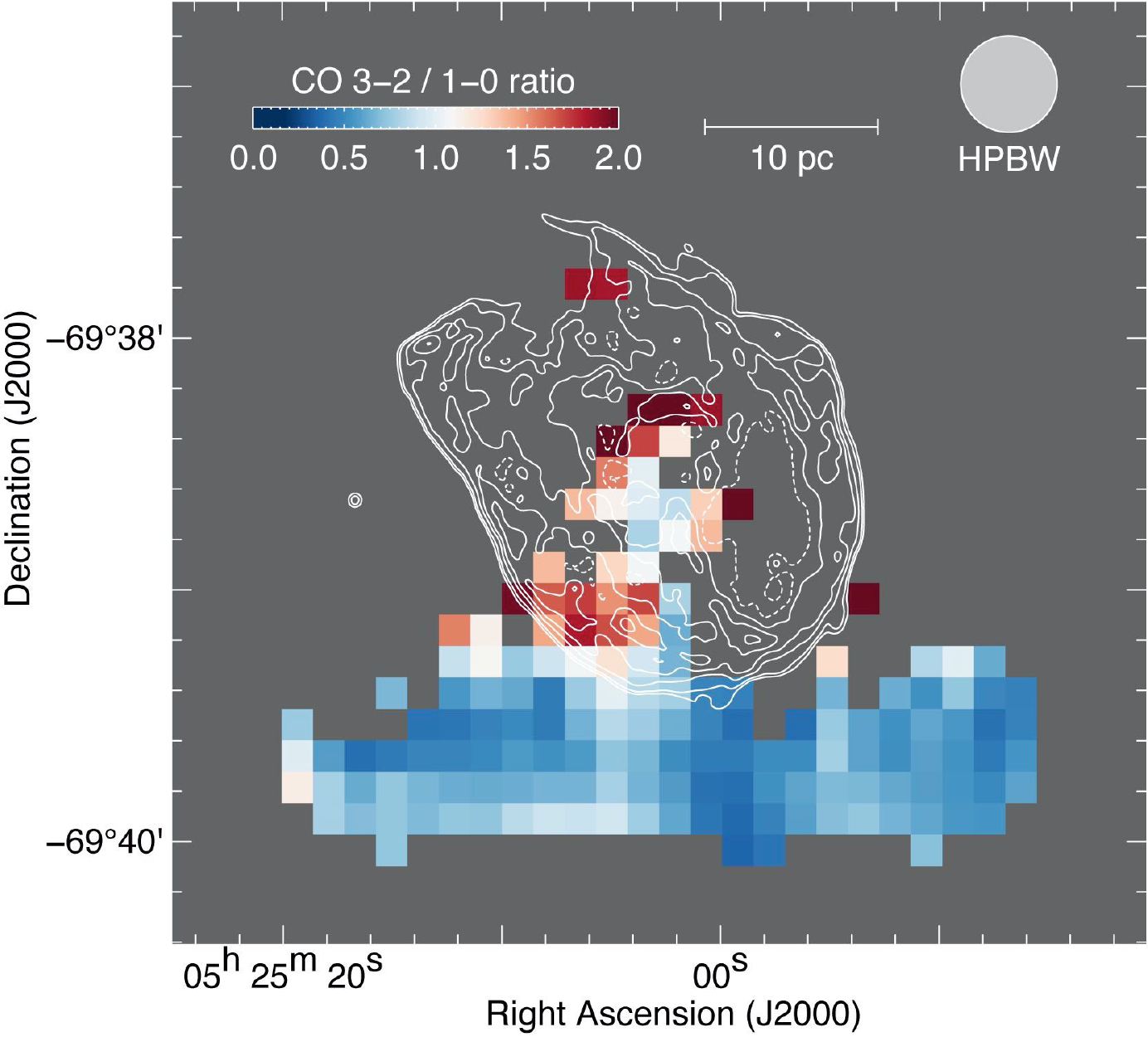}
\caption{Intensity ratio map of $^{12}$CO($J$ = 3--2) / $^{12}$CO($J$ = 1--0) using ASTE and ALMA. The ALMA data was smoothed to match the effective beam size of the ASTE data (an effective beam size of $23''$). The beam size and scale bar are also shown in top right corner. The velocity range is same as in Figure~\ref{fig5}. White dashed contours represent the X-ray intensity, whose contour levels are the same as in Figure~\ref{fig2}b. The gray areas represent that the $^{12}$CO($J$ = 1--0) and/or $^{12}$($J$ = 3--2) data show the low significance of $\sim$8$\sigma$ or lower.}
\label{fig6}
\end{center}
\end{figure}%

Figure~\ref{fig6} shows an intensity ratio map of $^{12}$CO($J$ = 3--2) / $^{12}$CO($J$ = 1--0) (hereafter $R_{3\text{--}2/1\text{--}0}$) using ASTE and ALMA, overlaid with {\it{Chandra}} X-ray contours. The intensity ratio reflects the CO rotational excitation states of the molecular clouds, and hence high intensity ratio $R_{3\text{--}2/1\text{--}0}$ indicates high temperature of the cloud. We find a high intensity ratio $R_{3\text{--}2/1\text{--}0}$ of $\sim$1.5--2.0 within the X-ray shell boundary. On the other hand, the intensity ratio $R_{3\text{--}2/1\text{--}0}$ of N132D GMC-south, south of the SNR, is $\sim$0.4, corresponding to the typical values of quiescent molecular clouds without any embedded OB association and/or shocks \citep[e.g.,][]{2019A&A...628A..96C}.

\subsection{Comparison with O-rich Ejecta}\label{ss:orich}

\begin{figure*}[]
\begin{center}
\includegraphics[width=\linewidth,clip]{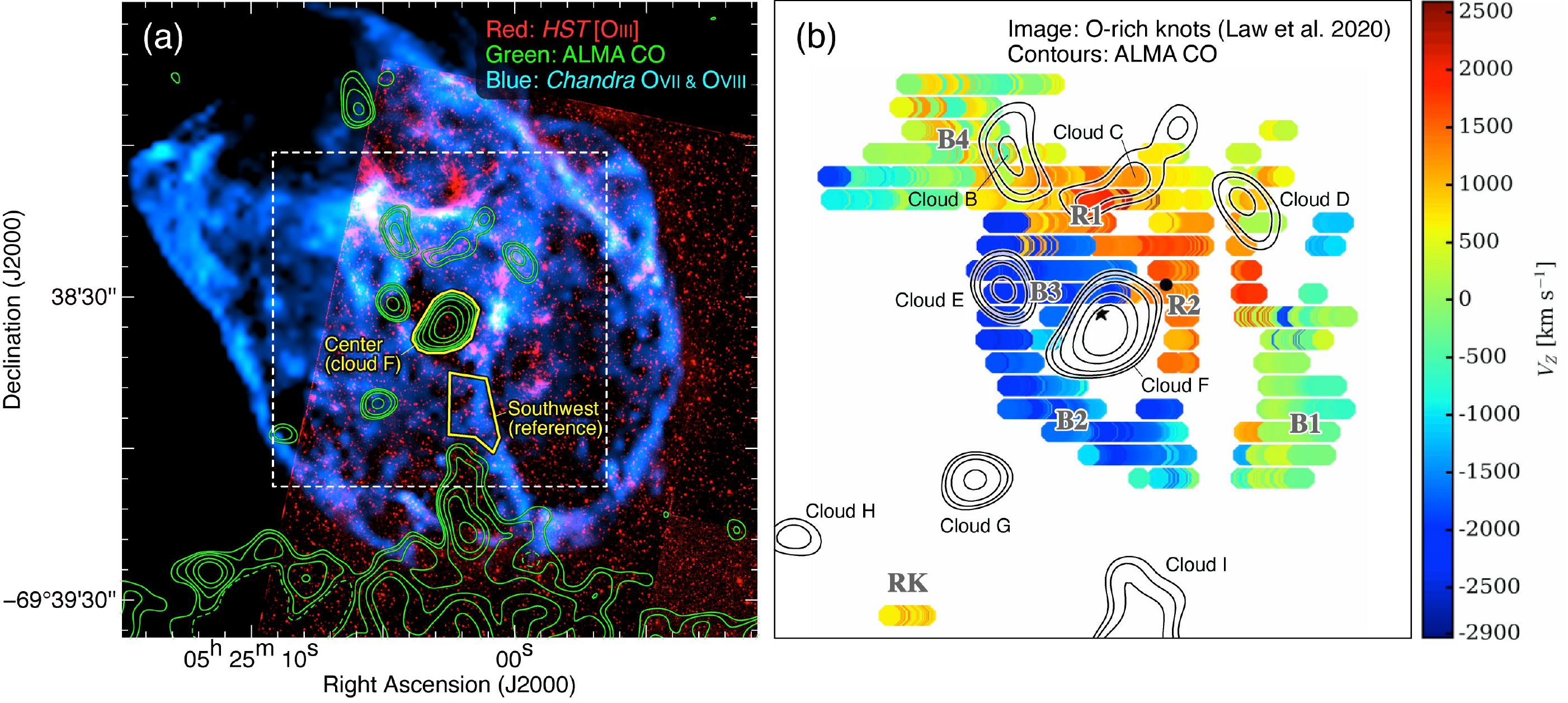}
\caption{(a) RGB image of N132D obtained with the {\it HST} [O {\sc iii}] (red), ALMA $^{12}$CO($J$ = 1--0) (green contours), and {\it Chandra} O {\sc viii} emission produced using a novel method for source separation (blue, see Appendix). The {integration velocity range and} contour {levels of CO} are the same as in Figure~\ref{fig5}. Regions enclosed by solid yellow lines are used for the X-ray spectral analysis (see Section \ref{ss:xspec}). The dashed white box represents the area shown in Figure \ref{fig7}(b). (b) Radial velocity distributions of O-rich knots presented by \cite{2020ApJ...894...73L}. Superposed contours indicate the ALMA $^{12}$CO($J$ = 1--0), whose contour levels are the same as in Figure~\ref{fig5}. The star and filled circle represent the center of O-rich knots and that of X-ray shell. Major knots --- B1, B2, B3, B4, R1, R2, and RK (runaway knot) --- from \cite{1995AJ....109.2104M} are also indicated.}
\label{fig7}
\end{center}
\end{figure*}%

Figure~\ref{fig7}a shows an enlarged view of the central region of N132D containing molecular clouds (green contours) and O-rich ejecta as seen by optical [O{\sc iii}] emission (red) and O{\sc vii} plus O{\sc viii} image of X-rays (blue). The optical [O{\sc iii}] emission is especially bright toward clouds B and C, also known as Lasker's Bowl \citep{1996AJ....112..509M}. The intercloud region between clouds D and F is also bright in both the [O{\sc iii}] and O{\sc vii} plus O{\sc viii} emission, whereas no bright O-rich ejecta is detected toward the center of cloud F. We find no apparent trend between the spatial distributions of the molecular clouds and O-rich ejecta. Figure~\ref{fig7}b shows radial velocity distributions of the O-rich knots presented by \cite{2020ApJ...894...73L}, overlaid with the ALMA CO contours. Major O-rich knots --- B1, B2, B3, B4, R1, R2, and RK (runaway knot) --- defined by \cite{1995AJ....109.2104M} are also indicated. We find that clouds B--F are projected onto the O-rich knots; Cloud E is in contact with the blue-shifted O-rich knots, whereas clouds B--D lie in the red-shifted O-rich knots. Cloud F is possibly associated with both the blue and red-shifted O-rich knots. It is noteworthy that cloud F shows a good spatial coincidence with the kinematic center of O-rich knots (marked as star symbol). Note that there are no CO counterparts of B1 and RK.

\subsection{X-ray Spectral Analysis}\label{ss:xspec}

\begin{figure*}[]
\begin{center}
\includegraphics[width=\linewidth,clip]{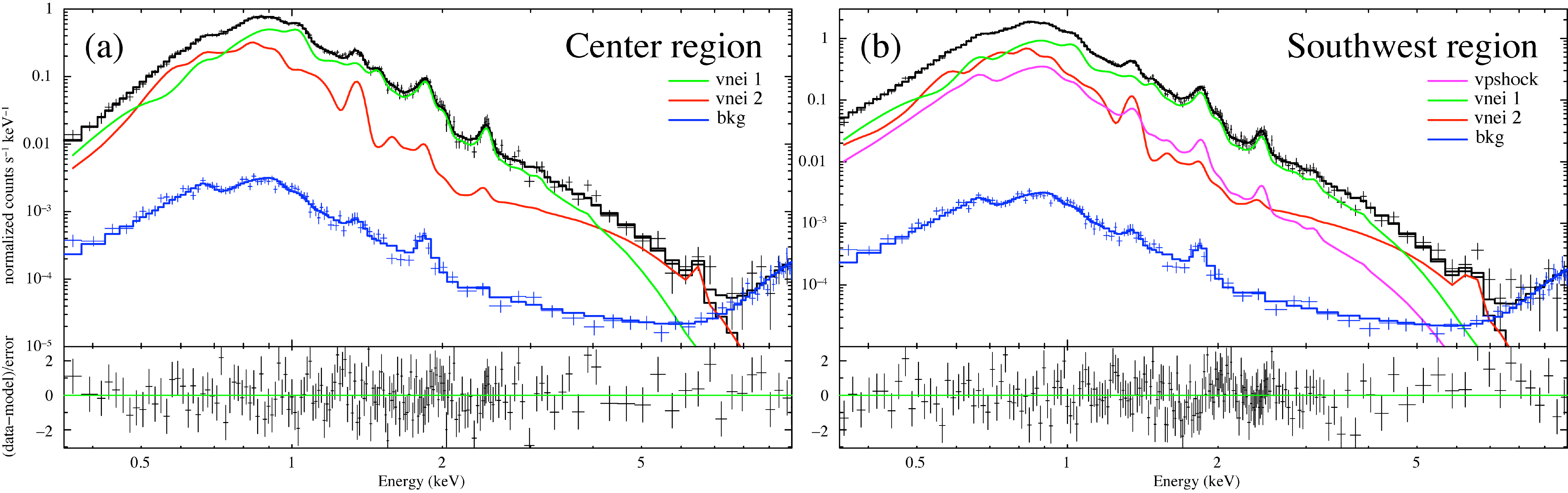}
\caption{(a) ACIS-S spectra of the center (cloud F) and southwest (reference), with the best-fit models shown in the top panels. The magenta, green, red, and blue lines represent {\tt vpshock}, {\tt vnei}~1, {\tt vnei}~2, and background components, respectively. The black lines represent the sum of the components. The bottom panels in each figure indicate (data $-$ model) / error.}
\label{fig8}
\end{center}
\end{figure*}%

To investigate the relationship between the molecular clouds and the X-ray emission of N132D, we derive the absorbing column densities toward two regions: one at the position of cloud F, and the other at a reference position south of cloud F (see Figure~\ref{fig7}a). The absorbing column density is useful to constrain the origin of the X-ray emission. This might provide evidence for a possible positional relationship between the cloud and the shock front \citep[][Y. Yamane et al. submitted to ApJ]{2015ApJ...799..175S,2019ApJ...873...40S}.

We extracted ACIS-S3 spectra from the regions labelled as the ``center region'' and ``southwest region'' in Figure~\ref{fig7}a. We extracted the background spectrum from two rectangular regions with a total area of $2.8~\mathrm{arcmin^2}$ to provide sufficient statistics. One region was located to the southwest of the remnant and the other to the northeast, both were positioned to include the contribution from the transfer streak of the CCD. We use the background model of \cite{Sharda2020} to fit both the source and the background spectra for each region.

We used a two-component absorption model comprised of the Milky Way absorption $N_\mathrm{H,MW}$ (TBabs) and the LMC absorption $N_\mathrm{H,LMC}$ (TBvarabs) by the ISM within the LMC along the line of sight. We fixed the hydrogen column density of the Milky Way at $1.47 \times 10^{21}$~cm$^{-2}$ \citep{2016AA...594A.116H} with solar abundance \citep{Wilms2000}. Further, we set the elemental abundance for the ISM in the LMC with He = 0.9 $Z_\odot$ and the other elements = 0.5 $Z_\odot$ on the \cite{Wilms2000} scale.

Following the latest X-ray study of N132D \citep{Sharda2020}, we fitted each source spectrum with a plane-parallel shock model \citep[{\tt vpshock}, see][]{2001ApJ...548..820B} plus two NEI components ({\tt vnei} $+$ {\tt vnei}). The thermal emission from the forward shock along the outer rim has been modeled by \cite{Sharda2020} with a {\tt vpshock} model with a temperature of $kT_\mathrm{e} = 0.86$~keV and an ionization timescale of $1.94\times10^{11}~\mathrm{cm^{-3}~s}$. We fix these parameters at these values in our fits and only allow the normalization to vary. The {\tt vpshock} component is intended to represent any emission from the forward shock that may contribute to our spectra along the line of sight. The two {\tt vnei} components are intended to represent emission from a shock/cloud interface and/or shock-heated ejecta.

\begin{deluxetable}{llcc}[]
\tablecaption{Best-fit X-ray Spectral Parameters
\label{tab:xspec}}
\tablehead{
\colhead{} &
\colhead{Parameter} &
\colhead{Center} &
\colhead{Southwest}
} 
\startdata
	$N_{\rm H}$ & $N_{\rm H, LMC}~(10^{21}~\rm cm^{-2})$ & 1.04 $_{-0.11}^{+0.18}$ & $\leq 0.13$ \\
	~ & $N_{\rm H, MW}~(10^{21}~\rm cm^{-2})$ & 1.47 (fixed) & 1.47 (fixed)\\
	vnei 1 & $kT_e~\rm (keV)$ & 0.82 $_{-0.02}^{+0.03}$ & 0.79 $_{-0.02}^{+0.04}$\\
	~ & $Z_{\rm O}~\rm (solar)$ & 1.18 $_{-0.48}^{+0.33}$ & 1.92 $_{-0.50}^{+0.29}$\\
	~ & $Z_{\rm Ne}~\rm (solar)$ & 2.36 $_{-0.27}^{+0.25}$ & 1.88 $_{-0.25}^{+0.16}$\\
	~ & $Z_{\rm Mg}~\rm (solar)$ & 0.94 $_{-0.13}^{+0.11}$ & 0.70 $_{-0.10}^{+0.12}$\\
	~ & $Z_{\rm Si}~\rm (solar)$ & 0.86 $_{-0.10}^{+0.08}$ & 0.80 $_{-0.09}^{+0.07}$\\
	~ & $Z_{\rm S}~\rm (solar)$ & 0.70 $_{-0.12}^{+0.11}$ & 0.65 $_{-0.06}^{+0.06}$\\
	~ & $Z_{\rm Fe}~\rm (solar)$ & 0.30 $_{-0.05}^{+0.04}$ & 0.28 $_{-0.07}^{+0.05}$\\
	~ & $n_et~\rm (10^{13}~cm^{-3}~s)$ & $\geq 3.50$ & $\geq 3.50$ \\
	~ & norm~($10^{-2}~\rm cm^{-5}$) & 8.27 $_{-0.76}^{+0.68}$ & 13.17 $_{-1.82}^{+1.12}$\\
	vnei 2 & $kT_e~\rm (keV)$ & 3.36 $_{-0.46}^{+0.47}$ & 2.44 $_{-0.33}^{+0.49}$\\
	~ & $Z_{\rm O}~\rm (solar)$ & 0.87 $_{-0.18}^{+0.33}$ & 0.64 $_{-0.14}^{+0.22}$\\
	~ & $Z_{\rm Ne}~\rm (solar)$ & 1.38 $_{-0.30}^{+0.25}$ & 1.70 $_{-0.61}^{+0.36}$\\
	~ & $Z_{\rm Mg}~\rm (solar)$ & 2.05 $_{-0.20}^{+0.51}$ & 2.58 $_{-0.62}^{+0.79}$\\
	~ & $Z_{\rm Fe}~\rm (solar)$ & 2.36 $_{-0.47}^{+1.21}$ & 6.38 $_{-1.94}^{+4.43}$\\
	~ & $n_et~\rm (10^{10}~cm^{-3}~s)$ & 0.66 $_{-0.06}^{+0.06}$ & 0.47 $_{-0.03}^{+0.01}$\\
	~ & norm~($10^{-3}~\rm cm^{-5}$) & 4.65 $_{-2.58}^{+1.16}$ & 6.64 $_{-2.47}^{+2.40}$\\
	vpshock & $kT_e~\rm (keV)$ & 0.86 (fixed) & 0.86 (fixed) \\
	~ & $n_et~\rm (10^{11}~cm^{-3}~s)$ & 1.94 (fixed) & 1.94 (fixed) \\
	~ & norm~($10^{-2}~\rm cm^{-5}$) & 0.0 & 3.03 $_{-0.02}^{+0.02}$\\
     \hline
    ~ & cstat (d.o.f.) & 1350 (1302) & 1427 (1302)\\
    ~ & Pearson-$\chi^2$ (reduced) & 1287 (0.99) & 1460 (1.12) \\
      \hline
\enddata
\end{deluxetable}

Figure~\ref{fig8} and Table \ref{tab:xspec} show the spectral fit results and the best-fit parameters, respectively. 
For the center region (cloud F), the C statistic is 1350 with 1302 degrees of freedom (DOF) and the Pearson reduced $\chi^2$ is 0.99. The normalization of the {\tt vpshock} component went to 0.0 and the LMC absorption went to $1.03 \pm 0.10 \times 10^{21}$~cm$^{-2}$. One {\tt vnei} component goes to a moderate temperature ($\sim$0.82 keV) and has the abundances of O, Ne, Mg, Si, S and Fe free to vary. There is marginal evidence for enhanced O, Ne, Mg, Si, and S abundances but the Fe abundance is consistent with LMC values.
The other {\tt vnei} goes to a high temperature (3.36~keV) and has O, Ne, Mg, and Fe free. 
The Ne, Mg, and Fe abundances are significantly enhanced compared to mean local LMC values.
Note that the ionization timescale for the 0.82~keV component goes to a value consistent with collisional ionization equilibrium (CIE; $3.5 \times 10^{13}$~cm$^{-3}$~s), while the ionization timescale for the 3.36~keV component goes to a low value of $6.6 \times 10^{9}$~cm$^{-3}$~s, indicating that the plasma producing this emission has been shocked relatively recently. For the southwest region (the reference region), the C statistic is 1427 with 1302 DOF and the Pearson reduced $\chi^2$ is 1.12. The {\tt vpshock} component is now a significant contributor (see magenta line in Figure~\ref{fig8}b). The LMC absorption is now 0.0, with an upper limit of $1.3 \times 10^{20}$~cm$^{-2}$. The fitted parameters for the two {\tt vnei} components for the center and southwest regions are similar to each other, most values are within $1.0\sigma$ of each other. The major difference is the absence of the {\tt vpshock} component in the center spectrum and the additional absorption for the center spectrum. This can be seen as the difference between the two spectra in the 0.35--1.0~keV energy range. The southwest spectrum has more emission at these lowest energies than the center spectrum and this emission is modeled by the {\tt vpshock} component. This indicates that there is additional absorption along the line of sight to the center region than the one toward the southwest region.

\subsection{Comparison with hard X-ray emission}\label{ss:HardXray}

\begin{figure}[]
\begin{center}
\includegraphics[width=\linewidth,clip]{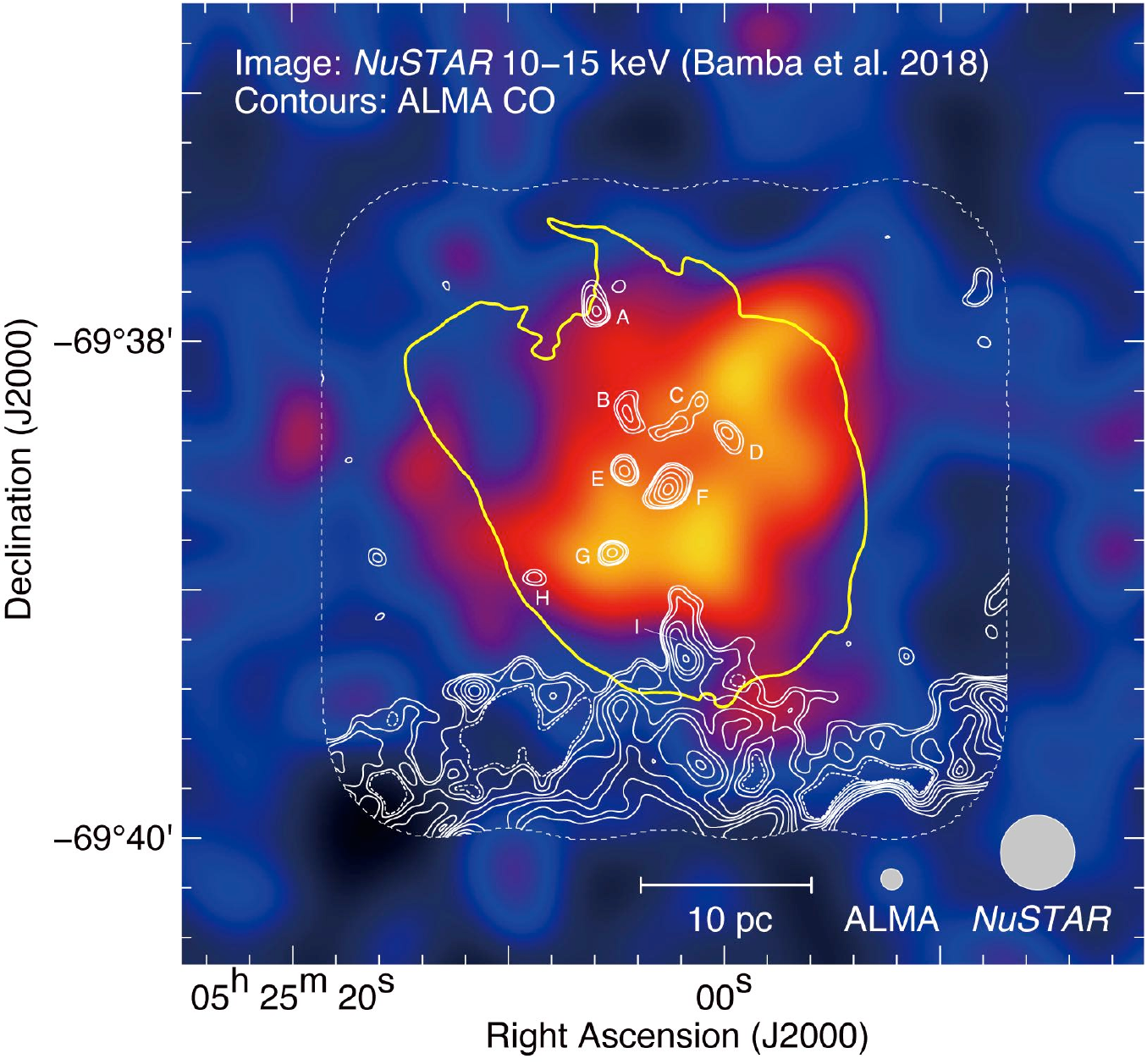}
\caption{Map of hard X-ray emission ($E$: 10--15 keV) obtained with {\it NuSTAR} \citep{2018ApJ...854...71B}. The white and yellow contours indicate the ALMA $^{12}$CO($J$ = 1--0) and the boundary of X-ray shell. The contour levels and integration velocity range of CO are the same as in Figure~\ref{fig5}. The beam size and scale bar are also shown in bottom right corner for each panel. The CO clouds A--I are also indicated.}
\label{fig9}
\end{center}
\end{figure}%

Figure~\ref{fig9} shows an overlay map of 10--15~keV band image obtained with {\it{NuSTAR}} \citep[colored image,][]{2018ApJ...854...71B} and the ALMA CO distribution in contours. 
The 10--15~keV band image represents possible synchrotron X-ray emission,
although \citet{2018ApJ...854...71B} could not exclude
the possibility of a very high temperature plasma emission.
We note that the hard X-ray emission are concentrated inside the SNR, where the molecular clouds B--G are located. Although the local intensity peaks of hard X-rays appear to be offset from the center of the molecular clouds, it is not certain whether the trend is significant because of the modest angular resolution of {\it{NuSTAR}} $\sim18''$ in HPBW ($\sim$4.4~pc at the LMC distance). It is certain that the edge of N132D is not bright with the hard X-ray emission, which is not typical for young SNRs with synchrotron X-rays
\citep{2005ApJ...621..793B}.

\section{Discussion}\label{s:discussion}

\subsection{Molecular Clouds associated with N132D}\label{ss:discussion1}
In addition to the previously known GMC, which we refer to as N132D GMC-south, we identified eight new molecular clouds toward the center and southern edge of N132D. To better understand the relationship among the clouds, high-energy radiation, and O-rich ejecta in N132D, it is essential to know which clouds are physically associated with the SNR. Here, we argue that the eight new molecular clouds resolved by ALMA are likely interacting with shockwaves and lie inside a wind-blown bubble.

We first claim that the high-intensity ratio of $R_{3\text{--}2/1\text{--}0} > 1.5$--2.0 as shown in Figure~\ref{fig6} provides strong evidence for shock-cloud interaction. The ratio of $R_{3\text{--}2/1\text{--}0}$ is useful to measure the degree of rotational excitation of CO molecules, because the upper state of $J$ = 3 lies at 33.2 ~K from the ground state of $J$ = 0, corresponding to $\sim$28~K above the state of $J$ = 1 at 5.5~K. The higher ratio of $R_{3\text{--}2/1\text{--}0}$ can trace warm molecular clouds heated by shock interactions not only for the Galactic SNRs (e.g., W28, \citeauthor{1999PASJ...51L...7A} \citeyear{1999PASJ...51L...7A}; Kesteven~79, \citeauthor{2018ApJ...864..161K} \citeyear{2018ApJ...864..161K}), but also for the Magellanic SNRs (e.g., LMC SNR N49, \citeauthor{2018ApJ...863...55Y} \citeyear{2018ApJ...863...55Y}; SMC SNR RX~J0046.5$-$7308, \citeauthor{2019ApJ...881...85S} \citeyear{2019ApJ...881...85S}). It is noteworthy that pre-shocked gas in N132D GMC-south shows significantly lower intensity of $R_{3\text{--}2/1\text{--}0} \sim$0.4, which is typical ratios of a quiescent cloud in the LMC without external heating \citep[e.g.,][]{2019A&A...628A..96C}.

\begin{figure}[]
\begin{center}
\includegraphics[width=\linewidth,clip]{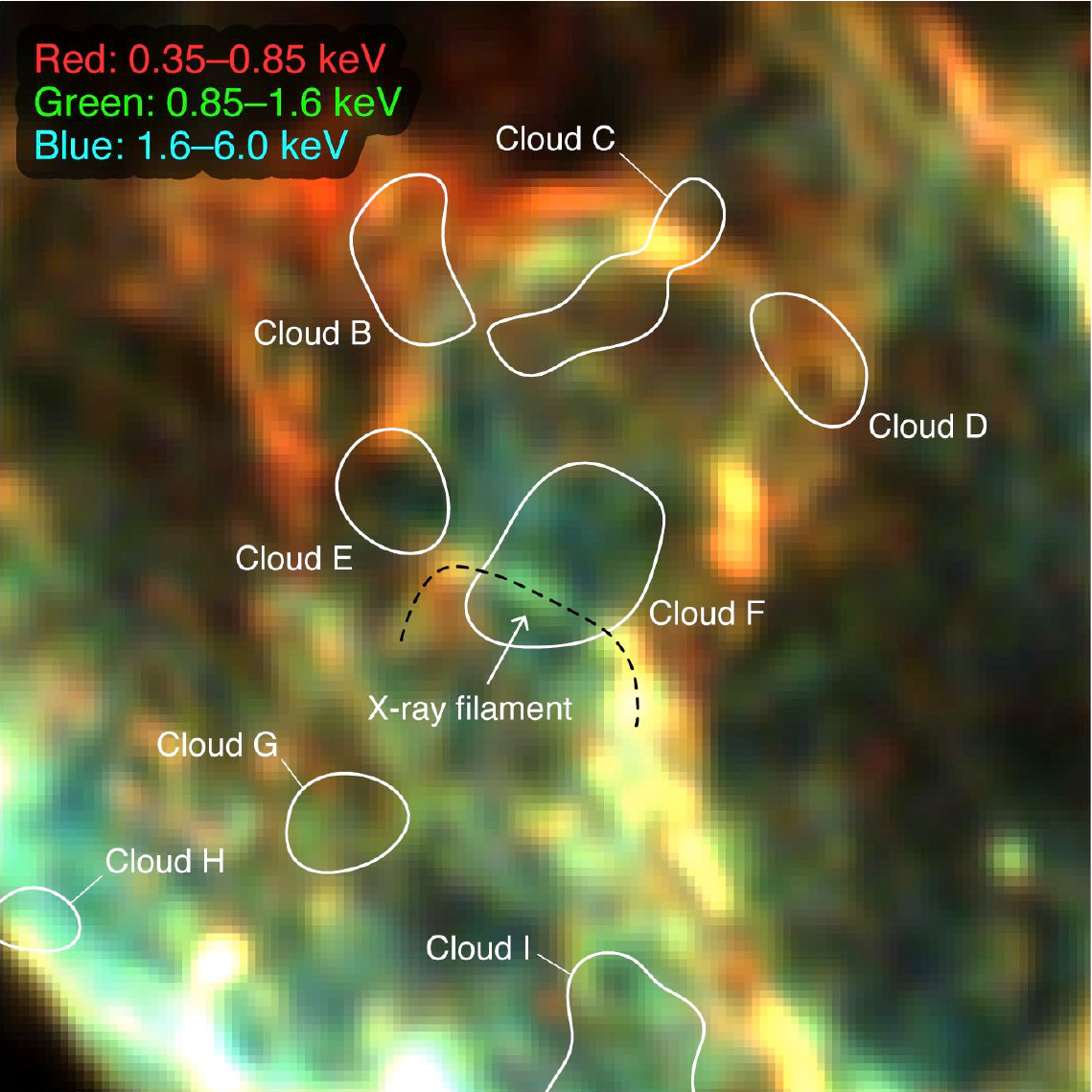}
\caption{RGB X-ray image toward the center of N132D. The red, green, and blue colors represent the energy bands of 0.35--0.85~keV, 0.85--1.6~keV, and 1.6{--}6.0~keV, respectively. Superposed contours indicate boundaries of ALMA CO clouds {B}--I as shown in Figure \ref{fig5}, whose contour levels are 1.2 K km s$^{-1}$. {Note that cloud A is located outside the field of view.} The dashed line indicates the X-ray filament, discussed in Section \ref{ss:discussion1}.}
\label{figex}
\end{center}
\end{figure}

We argue that cloud F has been completely engulfed by shocks and is located on the near side of remnant. Figure \ref{figex} shows an enlarged view of the X-ray three-color image superposed on boundaries of molecular clouds. We find an X-ray filament toward cloud F. The color changes from red/yellow to green as one moves from East to West onto the cloud. This indicates that low energy X-rays are suppressed toward cloud F due to absorption. In fact, the LMC absorption $N_\mathrm{H,LMC}$ of cloud F ($1.04 \times 10^{21}$ cm$^{-2}$) is significantly higher than that of the reference region without dense clouds ($\leq0.13 \times 10^{21}$ cm$^{-2}$). In this case, the forward shock likely propagated from behind cloud F to in front of it. Then, the X-ray filament was formed behind cloud F via shock interaction. This interpretation is also consistent with the absence of the {\tt vpshock} component. If the shock wave is in the process of wrapping around the cloud, the thermal emission from the forward shock would be suppressed on the near side of the cloud as the shock reforms on that side of the cloud. In addition, any thermal emission from the forward shock on the far side of the remnant is absorbed by the cloud. Both effects lead to a reduction in the thermal emission located along the line of sight to the center of the cloud.

The fitted LMC absorption $N_\mathrm{H,LMC}$ of $\sim1 \times 10^{21}$ cm$^{-2}$ toward cloud F also suggests that the molecular cloud is engulfed by shockwaves. By using equation (\ref{eq1}), we can derive an average proton column density of cloud F to be $\sim5 \times 10^{21}$~cm$^{-2}$, which is five times higher than the X-ray derived value (see Section~\ref{ss:xspec} and Table~\ref{tab:xspec}). This suggests that some of the X-ray emission originates in front of the cloud and not just behind it. The evaporating cloud scenario described in \cite{1977ApJ...211..135C} and \cite{1991ApJ...373..543W} and further explored by \cite{2019MNRAS.482.1602Z} can produce such a morphology. A similar discussion is also applicable for cloud I in the southern edge of the SNR. According to \cite{Sharda2020}, the LMC absorption $N_\mathrm{H,LMC}$ toward cloud I is $\sim$2--$3 \times 10^{21}$~cm$^{-2}$, two times lower than the average proton column density of cloud I. Although cloud I is located on the shell boundary of the SNR, it is likely that the shock is interacting with this cloud.

Additionally, the large $n_\mathrm{e}t$ value of the CIE plasma in cloud F is possibly consistent with a long-elapsed time since the cloud was heated. Our ALMA observations revealed the molecular hydrogen density of cloud F to be 930~cm$^{-3}$. Considering the post-shocked gas density equals to 1/4 of the pre-shock gas density in the limit of large Mach number (c.f., Rankin-Hugnoiot shock jump conditions), the electron density toward the region can be derived to $\sim$560~cm$^{-3}$ assuming the electron-to-proton density ratio of 1.2. We then obtain the ionization time of $\ga 2000$ yr, which is roughly consistent with the latest estimation of the SNR age of $2450 \pm 195$~yr \citep{2020ApJ...894...73L}. We therefore propose a possible scenario that cloud F is completely engulfed by shocks soon after the supernova explosion.

We also argue that the molecular clouds we observe through ALMA were left behind inside a wind-blown bubble. According to \cite{2012ApJ...744...71I}, the surrounding ISM of a high-mass progenitor shows a highly inhomogeneous density distribution. The less dense gas such as \HI\ clouds can be completely disrupted by the strong stellar winds, while the more dense gas such as molecular clouds can survive. As a result, a wind-blown bubble with a density of $\sim$0.01~cm$^{-3}$ coexists with dense molecular clouds with density more than $\sim10^3$~cm$^{-3}$. For N132D, the wind cavity and dense clouds are seen in Figures~\ref{fig2}c and \ref{fig3}. The \HI\ cloud shows a cavity-like distribution in both the spatial and velocity planes. The expansion velocity of $\sim$6~km~s$^{-1}$ is consistent with the typical gas motion seen in other core-collapse SNRs \citep[e.g.,][]{2012ApJ...746...82F,2018ApJ...864..161K,2019ApJ...881...85S}. We note that the wind-bubble explosion scenario is also proposed by previous optical and X-ray studies of N132D \citep[][]{1987ApJ...314..103H,1995ApJ...439..365S,2000ApJ...537..667B,Sharda2020} and modeled by \cite{2003ApJ...595..227C}. Additionally, the estimated progenitor mass for N132D from the model of \cite{2013ApJ...769L..16C} for an SNR evolving in a cavity is in good agreement with that from nucleosynthesis modeling of ejecta-rich regions \citep{Sharda2020}.

We emphasize that although the shock-cloud interaction has occurred, most of the dense molecular clouds can survive the shock-erosion. When shock waves hit the dense clouds, the penetrating velocity can be described as $V_\mathrm{sh} / \sqrt{(n / n_0)}$, where $V_\mathrm{sh}$ is the shock velocity before collision, $n_0$ is the ambient density inside the wind-bubble ($n_0 \sim 0.01$ cm$^{-3}$), and $n$ is the number density of molecular cloud ($n \sim 930$ cm$^{-3}$ for the case of cloud F). Therefore, the shock waves in cloud F will be much decelerated to $1/300$~$V_\mathrm{sh}$, and hence the shock cannot penetrate into the cloud within a few thousand years. The numerical results also support this idea \citep[e.g.,][]{2019MNRAS.487.3199C}. Furthermore, evaporation of the shocked cloud is also negligible due to the small thermal capacity of SNR's shocks. \cite{1990ApJ...351..157T} and \cite{2006ApJ...642L.149S} discovered shocked molecular clouds in the middle-aged Galactic SNR G109.1$-$1.0 with the age of $\sim1.4 \times 10^4$ yr \citep{2013A&A...552A..45S}, which survived the encounter. The surviving clouds, with a total mass of 63 $M_\odot$, are associated with a thermal X-ray lobe. The authors conclude that the X-ray lobe was likely formed by the evaporation of a small outer portion of the clouds; the mass ratio of the thermal plasma relative to the molecular clouds is less than 10\% \citep{2006ApJ...642L.149S}. This interpretation is also supported by numerical simulations \citep[][]{2015A&A...582A..47B}. For N132D, clouds B--F have been partially evaporated because the bright H$\alpha$ emission and thermal X-rays are seen in the vicinity of them (see Figures \ref{fig2} and \ref{fig5}). In other words, almost all part of the clouds in N132D likely survived shock erosion or evaporation, considering the large total cloud mass of $\sim$750 $M_\odot$ and young age of N132D ($\sim$2500 yr). Therefore N132D may be considered to represent an early stage of the mixed-morphology SNR \cite[e.g.,][]{1998ApJ...503L.167R}.

The hard X-ray enhancements around the shocked clouds also provides alternative evidence for an inhomogeneous density distribution of the ISM. When a supernova shockwave propagates into an inhomogeneous ISM with a density fluctuation of roughly $10^5$, the shock-cloud interaction may generate turbulence that enhances the magnetic field up to $\sim$1~mG at the surface of the shocked clouds \citep[e.g.,][]{2009ApJ...695..825I,2012ApJ...744...71I}. This can be observed as synchrotron X-rays or a radio continuum enhancement around the shocked gas clouds \citep[e.g.,][]{2010ApJ...724...59S,2013ApJ...778...59S,2015ApJ...799..175S,2017ApJ...843...61S,2017JHEAp..15....1S,2018ApJ...863...55Y,2018PASJ...70...77O}. On the other hand, the shock-cloud interaction will develop multiple reflected shock structures which can heat the gas up to high temperatures \cite[e.g.,][]{2019ApJ...873...40S}. For N132D, the observational trend in Figure~\ref{fig9} --- spatial correspondence between the 10--15~keV X-rays and molecular clouds B--F --- shows possible evidence for the shock-cloud interaction with the magnetic field amplification and/or shock ionization. To confirm this idea, we need conclusive evidence of a synchrotron X-ray and/or high-temperature plasma enhancement around the molecular clouds with sufficient angular resolution. A deep exposure with {\it{Chandra}} offers the possibility of extracting the hard X-ray spectral component spatially coincident with the molecular clouds.

In conclusion, the eight new molecular clouds presented here are possibly located inside the wind-blown bubble formed by stellar winds from the progenitor of N132D, and these clouds are likely engulfed by supernova shock waves. On the other hand, the relationship between these clouds and the O-rich ejecta is still unknown from the current dataset (see Section \ref{ss:orich} and Figure~\ref{fig7}). It is possible that the O-rich emission of optical and X-rays were efficiently produced by reverse shocks due to the shock interaction with the dense molecular clouds \citep[e.g.,][]{2015Sci...347..526M}. Future ALMA observations with $\sim$0.1~pc resolution will allow us to compare spatial and kinematic distributions of the ISM/circumstellar medium (CSM) and ejecta.

\subsection{Is N132D the Energetic Accelerator of Cosmic-Ray Protons?}\label{ss:discussion2}
N132D is thought to be a promising candidate for a hadronic gamma-ray emitter because of its bright TeV gamma-rays and very weak or absent synchrotron X-ray emission \citep{2018ApJ...854...71B}. Although a detailed spatial comparison between the CO data and the gamma-ray emission could not be carried out, the presence of shocked molecular clouds provides support for the hadronic origin of gamma-rays in N132D. Assuming that this hypothesis is correct, we derive the total energy of accelerated cosmic-ray protons, $W_\mathrm{p}$, in N132D taking into account the target gas density. Previous studies measured the values of $W_\mathrm{p} \sim$10$^{50}$--$10^{51}$~erg using the X-ray-based or model-dependent gas density \citep{2015Sci...347..406H,2018ApJ...854...71B}. Here, we reconsider the total energy of cosmic-ray protons in N132D using the neutral gas density which is derived from radio observations. 

It should be emphasized that TeV gamma-rays are emitted from the direction of N132D itself, rather than from the surrounding GMCs and/or \HI\ cloud, even after taking into consideration the PSF of gamma-ray image (Figure~\ref{fig1}). This implies that the surrounding three to four GMCs and the southern \HI\ cloud do not significantly contribute to the gamma-ray emission via the hadronic process\footnote{Note that future gamma-ray observatories with high-angular resolution and high sensitivity have a potential to detect gamma-rays from the escaped cosmic-ray protons from the surrounding GMCs and \HI\ cloud. The Cherenkov Telescope Array (CTA) can test the presence of high-energy particles escaped from N132D.}. In the present study we therefore focus on the target gas density within the shell of the SNR.

We estimate the target proton density within a wind-blown bubble. As discussed in Section~\ref{ss:discussion1}, the molecular clouds are located inside the wind bubble. Since the intercloud density in the bubble is thought to to be significantly low \citep[$\sim$0.01~cm$^{-3}$, e.g.,][]{1977ApJ...218..377W}, the only plausible mechanism to produce gamma-ray emission concentrated in the center of the SNR is if the cosmic-ray protons interact with the molecular clouds. According to \cite{2015Sci...347..406H}, the total energy of cosmic-ray protons $W_\mathrm{p}$ can be described in case of the hadron-dominant model:
\begin{eqnarray}
W_\mathrm{p} \sim 10^{52} (n_\mathrm{H} / 1\; \mathrm{cm^{-3}})^{-1} \;\; \mathrm{(erg)}, 
\label{eq3}
\end{eqnarray}
where $n_\mathrm{H}$ is the number density of interstellar protons. Adopting proton densities of molecular clouds of 260--1980~cm$^{-3}$ (see Table \ref{tab:mc}), we then obtain $W_\mathrm{p} \sim$0.5--$3.8 \times 10^{49}$~erg. This is comparable to the values obtained for the Galactic gamma-ray SNRs \citep[$\sim$10$^{48}$--10$^{49}$ erg, e.g.,][]{2012ApJ...746...82F,2017ApJ...850...71F,2013ApJ...768..179Y,2014ApJ...788...94F,2018ApJ...864..161K,2019ApJ...876...37S}. Note that the derived value gives a conservative lower limit on the total energy of cosmic-ray protons, because the hadronic gamma-ray emission can be observed only toward the gas cloud even if cosmic-ray protons have an azimuthally isotropic distribution. In other words, there are cosmic-ray protons that do not interact with the molecular clouds and do not produce gamma rays, and the value of $W_\mathrm{p}$ should be slightly increased. In any case, N132D can be classified as a common accelerator of cosmic-ray protons in the Local Group of galaxies.

We also discuss an alternative case that the shock front of N132D has reached the cavity wall of the wind-blown bubble. In this case, atomic hydrogen gas within the wind shell acts as the target for cosmic-ray protons. The column density of atomic hydrogen $N_\mathrm{p}$(\HI) is calculated using the following equation \citep{1990ARA&A..28..215D};
\begin{eqnarray}
N_\mathrm{p}(\mathrm{H\,}{\textsc{i}}\xspace) = 1.823 \times 10^{18}  \cdot W(\mathrm{H\,}{\textsc{i}}\xspace) \;(\mathrm{cm}^{-2}),  
\label{eq4}
\end{eqnarray}
where $W$(\HI) is the integrated intensity of \HI\ in units of K~km~s$^{-1}$. Since $W$(\HI) toward N132D has large uncertainty due to the radio continuum absorption (see Figures \ref{fig3} and \ref{fig4}), we derive it by referring to the \HI\ intensity surrounding the shell. The typical value of $W$(\HI) near the shell is to be $\sim$500~K~km~s$^{-1}$ (see Figure~\ref{fig2}c), and hence the average column density of atomic hydrogen is derived to $\sim0.9 \times 10^{21}$~cm$^{-2}$. Considering the wind bubble expansion, the atomic hydrogen gas was swept up within the thick wind shell. We here assume that the diameter and thickness of the \HI\ wind shell are $\sim$25~pc and $\sim$5~pc, respectively. The former corresponds to an effective diameter of N132D, and the latter represents the typical thickness of wind shell surrounding a high-mass star or core-collapse SNR \citep[e.g.,][]{2006ApJ...642..307Y,2012ApJ...746...82F,2017ApJ...850...71F,2019ApJ...876...37S}. We finally obtain the atomic hydrogen density within the wind shell to be $\sim$30~cm$^{-3}$, corresponding to $W_\mathrm{p} \sim3 \times 10^{50}$~erg. This energy is significantly higher than the values which are seen in Galactic gamma-ray SNRs, and hence N132D might be an energetic accelerator of cosmic-ray protons as mentioned before \citep{2015Sci...347..406H,2018ApJ...854...71B}. To confirm the shock-interaction with the wind shell, further \HI\ observations are needed. The Australian Square Kilometre Array Pathfinder (ASKAP), MeerKAT and the Square Kilometre Array (SKA) will be able to spatially resolve the wind-blown bubble of \HI\ with fine angular resolution and high sensitivity.

\section{Conclusions}\label{s:conclusions}
We have presented new $^{12}$CO($J$ = 1--0, 3--2) observations toward the LMC SNR N132D using ALMA and ASTE. The primary conclusions are summarized as follows.
\begin{enumerate}
\item We have revealed the presence of diffuse CO emission inside the X-ray shell in addition to the previously known GMC at the southern edge of N132D. ALMA observations spatially resolved the diffuse CO emission into nine molecular clouds, whose sizes and masses are 1.2--2.2 pc and 30--240 $M_\odot$. High-intensity ratios of CO $J$ = 3--2 / 1--$0 > 1.5$ are seen toward the molecular clouds, indicating that shock-heating has occurred. The expansion \HI\ shell with an expanding velocity of $\sim$6~km~s$^{-1}$ is also found toward N132D.
\item Spatially resolved X-ray spectroscopy has revealed that the emission from the line of sight to cloud F can be well represented by a model with absorption in excess of the LMC absorption of $N_{\rm H}=1.04\times 10^{21}$~cm$^{-2}$ and two NEI thermal components (one of which approaches CIE conditions) and no thermal component for the forward shock emission. On the other hand, the fit to the X-ray spectrum of an adjacent region off of cloud F shows no additional absorption compared to the LMC value and requires a thermal component for the forward shock in addition to the two NEI components. The larger absorption and absence of a thermal component associated with the forward shock along the line of sight to cloud F suggests that cloud F has been engulfed by shocks and is located on the near side of remnant.
\item We propose that the molecular clouds existed in the wind-blown bubble of the progenitor before the SNe explosion. The large $n_\mathrm{e}t$ value of one component of the plasma along the line of sight to cloud F is consistent with an elapsed time of $\ga 2000$~yr since the cloud was heated. The inhomogeneous density distribution inside the bubble --- diffuse gas of $\sim$0.01~cm$^{-3}$ and dense clouds of $\sim$1000~cm$^{-3}$ --- is also consistent with synchrotron X-ray and/or high-temperature plasma enhancement around the shocked clouds through the magnetic field amplification and/or shock ionization. 
\item If the hadronic process is the dominant contributor to the gamma-ray emission, the shock-engulfed molecular clouds play a role as targets for cosmic-rays. We estimate the total energy of cosmic-ray protons accelerated in N132D to be $\sim$0.5--$3.8 \times 10^{49}$~erg as a conservative lower limit, which is roughly the same value as seen in Galactic gamma-ray SNRs. The total energy could be as high as $\sim 3 \times 10^{50}$~erg if the shock front has reached the edge of the wind-blown cavity and the wind-shell of \HI\ has become a primary target for cosmic-ray protons. If the latter case is correct, N132D might be a very energetic accelerator of cosmic rays in the Local Group of galaxies.
\end{enumerate}

\software{CASA \citep[v 5.4.0.:][]{2007ASPC..376..127M}, CIAO \citep[v 4.12:][]{2006SPIE.6270E..1VF}, CALDB \citep[v 4.9.1:][]{2007ChNew..14...33G}, Xspec \citep{1996ASPC..101...17A}, SAO ds9 \citep{2003ASPC..295...489}}.

\section*{ACKNOWLEDGMENTS}
This paper makes use of the following ALMA data: ADS/JAO.ALMA\#2013.1.01042.S. ALMA is a partnership of ESO (representing its member states), NSF (USA) and NINS (Japan), together with NRC (Canada), MOST and ASIAA (Taiwan), and KASI (Republic of Korea), in cooperation with the Republic of Chile. The Joint ALMA Observatory is operated by ESO, AUI/NRAO and NAOJ. The ASTE radio telescope is operated by NAOJ. The Mopra radio telescope is part of the Australia telescope and is funded by the Commonwealth of Australia for operation as a National Facility managed by the CSIRO. The University of New South Wales Mopra Spectrometer Digital Filter Bank used for these Mopra observations was provided with support from the Australian Research Council, together with the University of New South Wales, the University of Adelaide, University of Sydney, Monash University, and the CSIRO. The scientific results reported in this article are based on data obtained from the Chandra Data Archive (Obs ID: 5532, 7259, and 7266). This research has made use of software provided by the Chandra X-ray Center (CXC) in the application packages CIAO (v 4.10). Based on observations made with the NASA/ESA Hubble Space Telescope, and obtained from the Hubble Legacy Archive, which is a collaboration between the Space Telescope Science Institute (STScI/NASA), the Space Telescope European Coordinating Facility (ST-ECF/ESA) and the Canadian Astronomy Data Centre (CADC/NRC/CSA). This research made use of the SAO/NASA Astrophysics Data System (ADS) bibliographic services. This work was supported by JSPS KAKENHI Grant Numbers JP16K17664 (H. Sano), JP19K14758 (H. Sano), JP19H05075 (H. Sano), and JP19K03908 (A. Bamba). H. Sano is also supported by the ALMA Japan Research Grant of NAOJ Chile Observatory (grant no. NAOJ-ALMA-244). This work is supported in part by Shiseido Female Researcher Science Grant (A. Bamba). P. Sharda is supported by the Australian Government Research Training Program (AGRTP) Scholarship. C. Law acknowledges funding from the National Science Foundation Graduate Research Fellowship under Grant DGE1745303. K. Tokuda was supported by NAOJ ALMA Scientific Research Grant Number of 2016-03B. M. Sasaki acknowledges support by the Deutsche Forschungsgemeinschaft through the Heisenberg professor grants SA 2131/5-1 and 12-1.

\begin{figure}[]
\begin{center}
\includegraphics[width=\linewidth,clip]{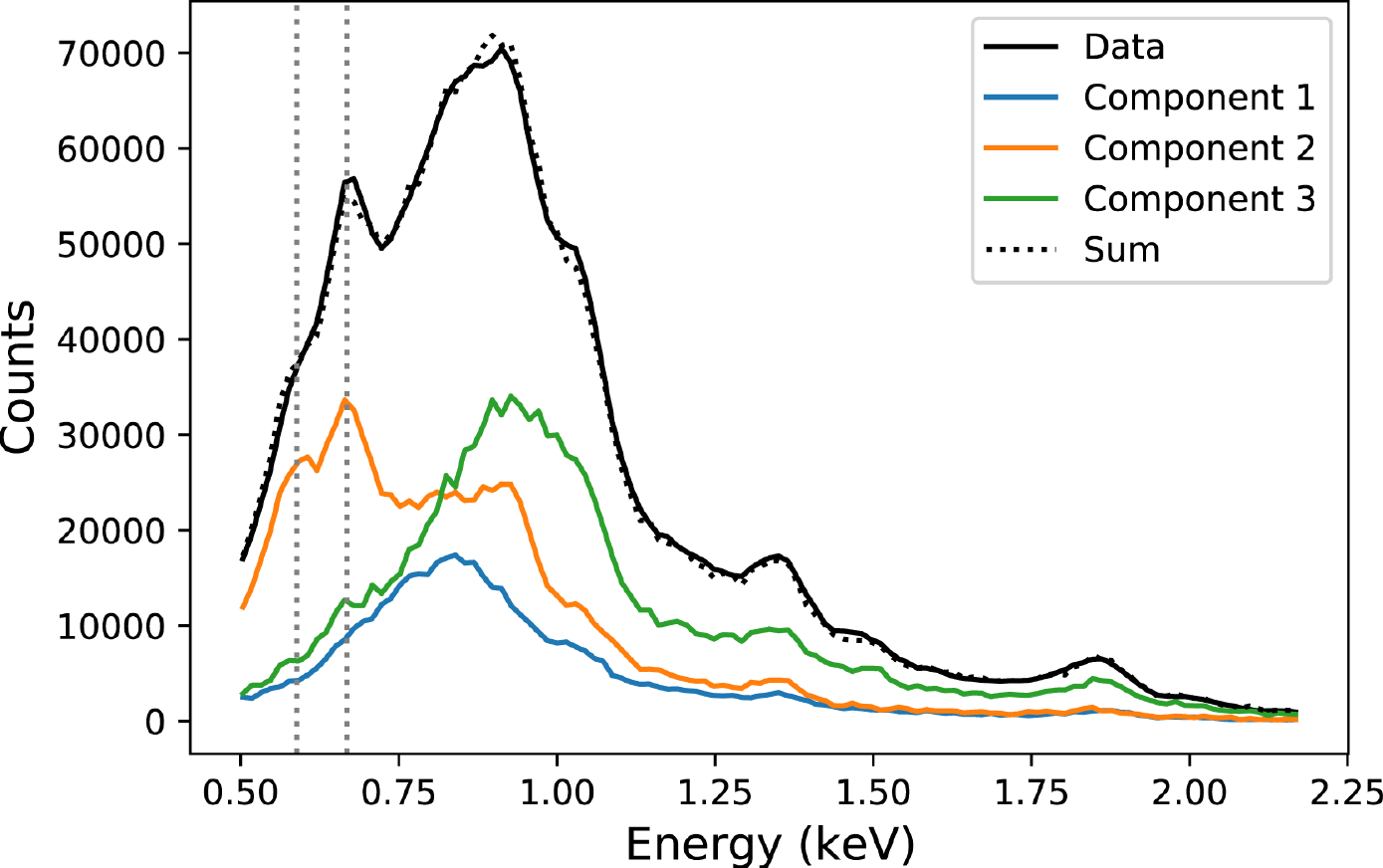}
\caption{Spectral decomposition of the X-ray data cube into different spectral components with a novel deblending method. The position of the O{\sc vii} and O{\sc viii} lines at 0.574 and 0.654 keV respectively are shown by the dotted lines.}
\label{gmca}
\end{center}
\end{figure}%

\section*{APPENDIX: An X-ray map for Oxygen dominated component}
Despite their multi-dimensional nature, X-ray data are most frequently analyzed as 2D images or 1D spectra independently therefore disconnecting the spatial and morphological information. To generate the oxygen X-ray map presented in Figure \ref{fig7}a we used a novel deblending technique recently adapted to X-ray data in \cite{2019A&A...627A.139P} that takes the full advantage of the 3D (X, Y, E) information provided by X-ray spectro-imagers. This method (the Generalized Morphological Component Analysis ; GMCA) was initially developed to separate the cosmic micro-wave background image from the foregrounds in {\it{Planck}} data \citep[][]{2015ITSP...63.1199B,2016A&A...591A..50B}. The general idea is to decompose the input X-ray data cube in a linear sum of images and associated spectra each component being different from the next one by its morphological and spectral signature. Note that the algorithm has a blind source separation approach and has no instrumental (instrument response) nor astrophysical (spectral emission) knowledge. Only the number of components to retrieve is fixed by the user. The main disentangling factor is the morphological diversity of each component in the wavelet domain\footnote{The wavelet transform is applied to each image slice of the data cube in order to enhance the contrast between small and large scale features.} and their associated spectral signatures.

Based on the same {\it{Chandra}} data set as presented in Section \ref{ss:xray}, a data cube was produced with the instrumental energy channel binning of 14.6 eV and a spatial bin size of 1.5 arcsec. The algorithm was applied in the 0.5 to 2.2 keV band and the number of components to retrieve was fixed to three. Figure \ref{gmca} shows the resulting spectral decomposition with one component dominating the low energies and exhibiting notable line emission at 0.574, 0.654 keV (dotted lines in Figure \ref{gmca}). Due to this spectral feature and the morphological similarities to the {\it{HST}} [O{\sc iii}] map (see Figure \ref{fig7}a), we associate this component with an oxygen rich component. The image associated with this spectral component is shown in Figure \ref{fig7}a (blue). As this component is dominating the 0.5--0.8 keV band, it is the most sensitive to absorption along the line of sight and the drops in flux in the image reflects the regions of highest absorption traced by the ALMA CO data.

\end{document}